\documentclass[11pt]{article}
\usepackage{epsfig,graphics,empheq, graphicx,amsmath,amssymb,amsfonts}
\usepackage{epstopdf}
\usepackage{color}
\usepackage{amsmath}
\usepackage{cases}
\usepackage[position=bottom]{subfig}
\usepackage{paralist}
\usepackage{enumitem}
\usepackage{graphicx}
\usepackage[percent]{overpic}
\usepackage{tikz}

\newsavebox{\largestimage}

\setlist{nolistsep}

\newcommand{\bm}[1]{\mbox{\boldmath{$#1$}}}

\def\u{{\bm u}}

\def\x{{\bm{x}}}

\def\0{{\bm 0}}

\def\cl {\nonumber \\}
\def\el {\nonumber}

\DeclareMathOperator*{\argmin}{arg\,min}

\begin{document}

\title{Coupling kinetic theory approaches for pedestrian dynamics and disease contagion 
in a confined environment}

\author{Daewa Kim$^*$ and Annalisa Quaini$^{**}$ \\
\footnotesize{
$^*$Department of Mathematics, West Virginia University, 94 Beechurst Ave, Morgantown, WV 26506}\\
\footnotesize{
$^{**}$Department of Mathematics, University of Houston, 3551 Cullen Blvd, Houston TX 77204}\\
\footnotesize{daewa.kim@mail.wvu.edu; quaini@math.uh.edu}
}

\maketitle

\noindent{\bf Abstract}
The goal of this work is to study an infectious disease spreading
in a medium size population occupying a confined environment. For this purpose, 
we consider a kinetic theory approach to model crowd dynamics 
in bounded domains and couple it to a kinetic equation to model contagion. 
The interactions of a person with other pedestrians and the environment 
are modeled by using tools of game theory. 
The pedestrian dynamics model allows to weight between two competing behaviors: the search for less congested 
areas and the tendency to follow the stream unconsciously in a panic situation. 
Each person in the system has a contagion level that is affected by the people in their neighborhood.
For the numerical solution of the coupled problem, we propose a numerical algorithm that 
at every time step solves one crowd dynamics problem and one contagion problem, i.e.~with no
subiterations between the two.
We test our coupled model on a problem involving a small crowd walking through a corridor. 

\section{Introduction} \label{sec:Intro}
We are interested in studying the early stage of an infectious disease spreading in intermediate size
populations who occupy confined environments. 
Models predicting the severity of an epidemic have all been based on
the averaged large population behaviors over a given calendar time. Such models are typically
compartmental models which use mean-field approximations, and are not valid when the size of the healthy population is 
small-to-medium, which is the case, e.g., in airports and schools. We propose to couple a kinetic theory approach to model
pedestrian dynamics, which has been successfully validated against experimental data for a medium-sized 
population in \cite{kim_quaini}, to a model for the evolution of the probability to find people 
with a given contagion level at a given position and time. We call the latter contagion model. 
We assume that the disease is such that it spreads with close proximity of individuals (like, e.g., measles, 
influenza etc.). Thus, the contagion model is coupled to the pedestrian dynamics model through people's position.

Because of our interest in confined environments and medium-sized crowds, we focus on a mesoscopic model for 
pedestrian dynamics. The scale of observation for the this approach is between the scale of 
macroscopic models (see, e.g., \cite{HUGHES2002507,5773492} and references therein)
and the scale of microscopic models (see, e.g.,
\cite{Chraibi2011425, BS:BS3830360405, Helbing1995, 1367-2630-1-1-313, 6701214, Moussad2755, TurnerPenn, 6248013} and references therein). 
Introduced in \cite{Bellomo2011383} and further developed in 
\cite{Agnelli2015,Bellomo2013_new,Bellomo2017_book,Bellomo2015_new,Bellomo2016_new,Bellomo2019_new,Bellomo2013},
the mesoscale approach derives a Boltzmann-type evolution equation for the statistical
distribution function of the position and velocity of the pedestrians,
in a framework close to that of the kinetic theory of gases.
See also \cite{Bellomo2012} for a literature review on this approach.
The model proposed in \cite{Bellomo2011383,Bellomo2013_new, Bellomo2013} is valid in
unbounded domains and with a homogeneous distribution of walking ability for the pedestrians, 
while the extension to bounded domains is presented in \cite{Agnelli2015}
and further explored in \cite{Bellomo2015_new,Bellomo2016_new,Bellomo2019_new}. 
In \cite{Bellomo2015_new}, more general features of behavioral-social dynamics are
taken into account. In \cite{Bellomo2016_new}, Monte Carlo simulations are introduced 
to study pedestrians behavior in complex scenarios. 
The methodology in \cite{Bellomo2016_new} is 
validated by comparing the computed and empirically observed fundamental density-velocity diagrams
and by checking that well known emerging properties are reproduced.
A kinetic theory approach for modeling pedestrian dynamics in presence of social phenomena,
such as the propagation of stress conditions, is presented in \cite{Bellomo2019_new}. 
Finally, we refer to \cite{Bellomo2017_book} for a thorough description of
how kinetic theory and evolutionary game theory can be used to understand the dynamics of living systems.
In \cite{kim_quaini}, we considered the model proposed in \cite{Agnelli2015} and extended 
it to bounded domains (rooms with one or more exits
with variable size) with obstacles. Moreover, we compared our numerical results against the data 
reported in a recent empirical study \cite{Kemloh}. We showed that 
for medium-sized groups of people the results produced by our model 
achieve good agreement between the computed average people
density and flow rate and the respective measured quantities.

In this work, we first consider a one dimensional version of the pedestrian dynamics model studied in 
\cite{Agnelli2015,kim_quaini} and then couple it to a one dimensional contagion model
inspired from the work on emotional contagion in \cite{Bertozzi2015}.
The model in \cite{Bertozzi2015} is adapted from the so-called 
ASCRIBE model \cite{Bertozzi2014ContagionSI,688c8c70f9c748e486b48806e7ce2fdb}, which has been designed to
track the level of fear within individuals under the assumption that it influences their motion. 
ASCRIBE has been implemented in agent-based simulation tool
ESCAPES \cite{tsai_escapes_2011} that has 
compared favorably to actual crowd footage with respect to other pedestrian models
\cite{inproceedings}. We modify the model in  \cite{Bertozzi2015} to account for disease spreading, 
instead of emotional contagion (i.e., spreading of fear or panic).
Since the pedestrian dynamics model and the contagion model are one-way coupled 
(the contagion is affected by people's position, while people's motion is independent
of the disease contagion), we propose a numerical algorithm that 
at every time step solves one crowd dynamics problem and one contagion problem, i.e.~with no
subiterations between the two.
We test our coupled model on a problem involving a small crowd walking through a corridor. 

For related work on coupled dynamics of virus infection and healthy cells
see, e.g., \cite{doi:10.1137/19M1250261} and references therein. 


The paper is organized as follows. 
Sec.~\ref{sec:ped_model} describes the modeling of one dimensional pedestrian dynamics
and presents results for a simple test case. 
In Sec.~\ref{sec:emotional}, we introduce the contagion model and validate it against the numerical 
results in \cite{Bertozzi2015}. The coupled model for pedestrian dynamics and disease
contagion and related numerical results are presented in Sec.~\ref{sec:coupling}.
Conclusions are drawn in Sec.~\ref{sec:concl}.

\section{One dimensional model for pedestrian dynamics}\label{sec:ped_model}

The model we consider is a 1D version of the model proposed in \cite{Agnelli2015}.
Let  $\Omega \subset \mathbb{R}$ denote a domain of inetrest.
We assume that pedestrians head to the exit $E$ within the domain.
Let $x\in \Omega$ denote position and $v_s=v \cos \theta$ denote
velocity, where $v$ is the velocity modulus and $\theta$ is the velocity direction (i.e.~either $\theta = 0$ or $\theta = \pi$).
For a system composed by a large number of pedestrians distributed inside
$\Omega$, the distribution function is given by
\[ f= f(t, x, v_s)\quad \text{for all} \,\,\, t \ge 0,  \,\, x \in \Omega. \]
Since we use polar coordinates for the velocity, we can write the distribution function as $f= f(t, x, v, \theta)$. 

We model the velocity magnitude $v$ as a continuous deterministic variable 
which evolves in time and space according to macroscopic effects determined by the overall dynamics.
This choice is supported by experimental evidence \cite{Schadschneider2011545}, which 
shows that in practical situations the speed of pedestrians depends mainly on the level of congestion around
them.
Due to the deterministic nature of the variable $v$, the kinetic type representation is given by the reduced distribution function
\begin{equation}\label{eq:f}
f(t, x, \theta)= \sum_{i=1}^{2} f^i(t, x)\delta(\theta - \theta_i),~\text{where}~\theta_{i}= (i-1)\pi,~\text{for}~i = 1, 2.
\end{equation}
In \eqref{eq:f}, $f^i(t, x)=f(t, x, \theta_i)$ represents the people that, at time $t$  and position $x$, move with direction $\theta_i$, while $\delta$ denotes the Dirac delta function.

Let us introduce some reference quantities that will be used to make the variables dimensionless.
We define:
\begin{itemize}
\item[-] $D$: the largest distance a pedestrian has cover in domain $\Omega$ to reach an exit;
\item[-] $V_M$: the highest velocity modulus a pedestrian can reach in low density and optimal environmental conditions;
\item[-] $T$: a reference time given by $D/V_M${;}
\item[-] $\rho_M$: the maximal admissible number of pedestrians per unit length.
\end{itemize}
The dimensionless variables are then: position $\hat{x}=x/D$, time $\hat{t}=t/T$, velocity modulus $\hat{v}=v/V_M$ and distribution function $\hat{f}=f/ \rho_M$. From now on, all the variables will be
dimensionless and hats will be omitted to simplify notation.

Due to the normalization of $f$, and of each $f^i$, the dimensionless local density is obtained by summing the distribution functions over the set of directions:
\begin{align}\label{eq:rho}
\rho(t, x)=\sum_{i=1}^{2}f^i(t, x) .
\end{align}
As mentioned earlier, the velocity modulus depends formally on the local density, i.e. $v=v[\rho](t, x)$, 
where square brackets are used to denote that $v$ depends on $\rho$ in a functional way.
For instance, $v$ can depend on $\rho$ and on its gradient.

A parameter $\alpha \in [0, 1]$ is introduced to represent the quality of the domain where pedestrians move:
$\alpha=0$ corresponds to the worst quality which forces pedestrians to slow down or stop,
while $\alpha=1$ corresponds to the best quality, which allows pedestrians to walk at the desired speed.
We assume that the maximum dimensionless speed $v_M$ a pedestrian can reach depends linearly on the quality
of the environment. For simplicity, we take $v_M = \alpha$. Let $\rho_c$ be a critical density
value such that for $\rho < \rho_c$ we have free flow regime (i.e.,~low density condition), while for $\rho > \rho_c$
we have a slowdown zone (i.e.,~high density condition). We set $\rho_c = \alpha/5$. Note that this choice is
compatible with the experimentally measured values of $\rho_c$ reported in \cite{Schadschneider2011545}.
Next, we set the velocity magnitude $v$ equal to $v_M$ in the free flow regime
and equal to a heuristic third-order polynomial in the slowdown zone:
\begin{align}\label{eq:v}
v=v(\rho)=
\begin{cases}
\alpha \quad & \text{for} \quad \rho \leq \rho_c(\alpha)= \alpha/5,  \\
a_3\rho^3+a_2\rho^2+a_1\rho+a_0 \quad & \text{for} \quad \rho > \rho_c(\alpha)=\alpha/5   ,
\end{cases}
\end{align}
where $a_i$ is constant for $i = 0,1,2,3$.
To set the value of these constants, we impose the following
conditions: $v(\rho_c) = v_M$,  $\partial_{\rho} v(\rho_c) = 0$, $v(1)=0$ and $\partial_{\rho} v(1) = 0$.
This leads to:
\begin{align}\label{eq:coeff}
\begin{cases}
a_0 &= (1/(\alpha^3-15\alpha^2+75\alpha-125))(75\alpha^2-125\alpha) \\
a_1 &= (1/(\alpha^3-15\alpha^2+75\alpha-125))(-150\alpha^2)\\
a_2 &= (1/(\alpha^3-15\alpha^2+75\alpha-125))(75\alpha^2+375\alpha) \\
a_3 &= (1/(\alpha^3-15\alpha^2+75\alpha-125))(-250\alpha).
\end{cases}
\end{align}

Finally, for direction $i = 1,2$ we set the velocity $v^i$
\begin{align}
v^i = v[\rho] \cos \theta_i. \label{eq:v^i} 
\end{align}

\subsection{Modeling interactions} \label{modelinginteractions}
Each pedestrian is modeled as a particle. Interactions involve three types of particles:

\begin{itemize}
\item[-] \textit{test particles} with state $(x, \theta_i)$: they are representative of the whole system;
\item[-] \textit{candidate particles} with state $(x, \theta_h)$: they can reach in probability the state of the test particles after individual-based interactions with the environment or with field particles;
\item[-] \textit{field particles} with state $(x, \theta_k)$: their presence triggers the interactions of the candidate particles.
\end{itemize}
A candidate particle modifies its state, in probability, into that of the test particle due to interactions with field particles,
while the test particle loses its state as a result of these interactions. 
The process through which a pedestrian decides the direction to take is the results of several factors.
We focus on three factors:

\begin{enumerate}[label={(F\arabic*})]
\item \textit{The goal to reach the exit.}\\
Given a candidate particle at the point $x$, we define its distance to the exit as
\[d_E(x)= \min_{x_E \in E} | x-x_E |,\]
and we consider the unit vector $\u_E(x)$, pointing from $x$ to the exit. Notice that vector $\u_E(x)$ is either
$(1,0)$ or $(-1,0)$.

\item \textit{The tendency to look for less congested area.}\\
A candidate particle $(x, \theta_h)$ may decide to change direction in order to avoid congested areas.
This is achieved with the direction that gives the minimal directional derivative of the density at the point $x$.
We denote such direction by unit vector $\u_C(\theta_h, \rho)$, which again is either
$(1,0)$ or $(-1,0)$.

\item \textit{The tendency to follow the stream.}\\
A candidate particle $h$ interacting with a field particle $k$  may decide to follow it and thus adopt its direction, denoted with unit vector $\u_F=(\cos\theta_k, 0)$.
\end{enumerate}

Factor (F1) is related to geometric aspects of the domain, while factors (F2) and (F3)
consider that people's behavior is strongly affected by the surrounding crowd.
Note that the effects related to factors (F2) and (F3) compete with each other: (F3) is dominant
in a panic situation, while (F2) characterizes rational behavior. As a weight between (F2) and (F3),
 we introduce parameter $\varepsilon \in [0,1]$: $\varepsilon=0$ corresponds to rational behavior,
 while $\varepsilon=1$ corresponds to panic behavior.

\subsubsection{The goal to reach the exit}
The goal to reach the exit, i.e.~the above (F1), is modeled with a term that involves: 
\begin{compactitem}
\item[-] \textit{transition probability} $\mathcal{A}_h(i)$: this is the probability that a candidate particle $h$, i.e.~with direction  $\theta_h$, adjusts its direction into that of the test particle $\theta_i$ due to the presence of an exit.
The following constraint for $\mathcal{A}_h(i)$ has to be satisfied:
\[
\sum_{i=1}^{2} \mathcal{A}_h(i)=1 \quad \text{for all} \,\, h \in \{1, 2\}.
\]
\item[-] \textit{rate} $\mu[\rho]$: this is the rate at which a candidate particle adjusts its direction as mentioned above. If the local density is getting lower, it is easier for pedestrians to see the exit. Thus, we set $\mu[\rho] =1-\rho$.
\end{compactitem}

We note that particles can, in probability, remain in state $h$ or
take the opposite direction. We define the vector
\begin{align}\label{eq:uG}
\u_G(\x, \theta_h) = (\cos \theta_G, 0).
\end{align}
Here $\theta_G$, which is either $\theta_1$ or $\theta_2$, is the  \textit{geometrical preferred direction}, which is the direction that a pedestrian should take in order to reach 
the exit in an optimal way. 
 
The transition probability for a candidate particle $h$ to switch its direction
with $\theta_G$ is given by:
\begin{equation}\label{eq:A}
\mathcal{A}_h(i)=\beta_h(\alpha)\delta_{s,i} + (1-\beta_h(\alpha))\delta_{h, i}, \quad i=1,2,
\end{equation}
where
\[s=\argmin_{j \in \{1,2\}}\{d(\theta_G, \theta_j)\},\]
with
\begin{equation}\label{eq:distance}
 d(\theta_p, \theta_q)= |\theta_p - \theta_q| 
\end{equation}
In \eqref{eq:A}, $\delta$ denotes the Kronecker delta function. Coefficient $\beta_h$ is defined by: 
\[
\beta_h(\alpha)=
\begin{cases}
\alpha & \text{if} \,\,\, d(\theta_h, \theta_G) = \Delta\theta, \\
0 & \text{if} \,\,\, d(\theta_h, \theta_G)= 0 ,
\end{cases}
\]
where $\Delta\theta=\pi$. 
Notice that if $\theta_G=\theta_h$, then $\beta_h=0$ and $\mathcal{A}_h(h)=1$, meaning that a
pedestrian keeps the same direction (in the absence of other interactions) with probability 1.

\subsubsection{Interactions between pedestrians}
To model the interaction with other pedestrians, we introduce:
\begin{itemize}
\item[-] \textit{transition probability} $\mathcal{B}_{hk}(i)[\rho]$: it gives the probability that a candidate particle
$h$ modifies its direction $\theta_h$ into that of the test particle $i$, i.e. $\theta_i$, due to the research
of less congested areas and the interaction with a field particle $k$ that moves with direction $\theta_k$.
The following constrain for $\mathcal{B}_{hk}(i)$ has to be satisfied:
\[
\sum_{i=1}^{2} \mathcal{B}_{hk}(i)[\rho]=1 \quad \text{for all} \,\, h, k  \in \{1, 2\},
\]
where again the square brackets denote the dependence on the density $\rho$.
\item[-] \textit{interaction rate} $\eta[\rho]$: it defines the number of binary encounters per unit time. If the local density increases, then the interaction rate also increases. For simplicity, we take $\eta[\rho]= \rho$.
\end{itemize}

The game consists in choosing the less congested direction among the two opposite directions. 
This direction can be computed for a candidate pedestrian $h$ situated at $x$, by taking
\[C=\argmin_{j \in \{1,2 \}}\{\partial_j\rho(t, x)\},\]
where $\partial_j\rho$ denotes the directional derivative of $\rho$ in the direction given by angle $\theta_j$.
We have $\u_C(\theta_h, \rho)=(\cos\theta_C, 0)$.
As for the tendency to follow the crowd, we set $\u_F=(\cos\theta_k, 0)$. This means that  a
candidate particle follows the direction of a field particle.

To take into account (F2) and (F3), we define the vector
\[\u_P(\theta_h, \theta_k, \rho)= \frac{\varepsilon\u_F+(1-\varepsilon)\u_C(\theta_h, \rho)}
{||\varepsilon\u_F+(1-\varepsilon)\u_C(\theta_h, \rho)||} = (\cos \theta_P, 0),
\]
where the subscript $P$ stands for \textit{pedestrians}. Direction $\theta_P$ is the \textit{interaction-based preferred direction}, obtained as a weighted combination between the tendency to follow the stream and the tendency to avoid crowded areas.

The transition probability is given by:
\[\mathcal{B}_{hk}(i)[\rho]=\beta_{hk}(\alpha)\rho\delta_{r, i} + (1-\beta_{hk}(\alpha)\rho)\delta_{h,i}, \quad i=1, 2,\]
where $r$ and $\beta_{hk}$ are defined by:
\[r=\argmin_{j \in \{1, 2 \}} \{d(\theta_P, \theta_j)\},\]

\[
\beta_{hk}(\alpha)=
\begin{cases}
\alpha & \text{if} \,\,\, d(\theta_h, \theta_P) = \Delta\theta, \\
0 & \text{if} \,\,\, d(\theta_h, \theta_P)=0.
\end{cases}
 \]


\subsection{One-dimensional equation of balance}\label{chap:1dkinetic}
The 1D mathematical model is obtained by a suitable balance of particles in an elementary area
of the space of microscopic states, considering the net flow into such area due to
transport and interactions \cite{Agnelli2015}.
Taking into account factors (F1)-(F3), we obtain:
\begin{align}
\frac{\partial f^i}{\partial t} &+ \frac{\partial \left( v^i [\rho] (t, x) f^i(t, x) \right)}{\partial x} \cl
& = \mathcal{J}^i[f](t, \x) \cl
& = \mu[\rho] \left( \sum_{h = 1}^2 \mathcal{A}_h(i) f^h(t, x) - f^i(t,x) \right) \cl
& \quad + \eta[\rho] \left( \sum_{h,k = 1}^2 \mathcal{B}_{hk}(i) [\rho] f^h(t, x)f^k(t, x) - f^i(t,x) \rho(t, x)\right)   
\label{eq:1dmodel}
\end{align}
for $i= 1,2$. In \eqref{eq:1dmodel}, functional $\mathcal{J}^i[f]$ models
the goal to reach the exit and the interaction with other people.

\subsection{The Lie operator-splitting scheme}\label{lie_scheme}

We apply the Lie operator-splitting scheme (see, e.g., \cite{glowinski2003finite}) to problem~\eqref{eq:1dmodel}. The problem 
will be split into two subproblems:
\begin{enumerate}
 \item A pure advection problem in the $x$ direction.
 \item A problem involving the goal to reach the exit and the interaction with other pedestrians.
\end{enumerate}

Let $\Delta t>0$ be a time discretization step for the time interval $[0, T]$. 
Denote $t^k=k\Delta t$, with $k = 0, \dots, N_t$ and let $g^k$
be an approximation of generic function $g$ at time $g^k$, i.e.~$g(t^k).$
Given an initial condition $f^{i,0}=f^i(0, x)$, for $i = 1, 2$, the Lie operator-splitting algorithm applied to to Problem~\eqref{eq:1dmodel} reads:
For $k=0,1,2, \dots, N_t-2,$ perform the following steps:
\begin{enumerate}
\item Find $f^i$, for $i = 1, 2$, such that
\begin{equation}
\begin{cases}
 \dfrac{\partial f^i}{\partial t} + \dfrac{\partial }{\partial x} \left( (v [\rho] \cos\theta_i) f^i(t, x) \right)=0, 
   \,\,\, \text{on } (t^k, t^{k+1}) \\ \label{eq:1dstep1}
f^i(t^k, x)=f^{i,k}
\end{cases} 
\end{equation}
 Set $f^{i,k+\frac{1}{2}}=f^i(t^{k+1}, x)$.
 
\bigskip
\item Find $f^i$, for $i = 1, 2$, such that\\
\begin{equation}
\begin{cases}
 \dfrac{\partial f^i}{\partial t} = \mathcal{J}^i[f](t, x)  \,\,\, \text{on } (t^k, t^{k+1})  \\ \label{eq:1dstep3}
f^i(t^k, x)=f^{i, k+\frac{1}{2}}
\end{cases}
\end{equation}
 Set $f^{i,k+1}=f^i(t^{k+1}, x)$.
\end{enumerate}
Once we compute $f^{i,k+1}$ for $i= 1, 2$, we use an equation \eqref{eq:rho}
to get the density $\rho^{k+1}$ and equations \eqref{eq:v}-\eqref{eq:coeff} to get the velocity magnitude at time $t^{k+1}$.

\subsection{Full discretization} \label{sec:1d-discretization}
We assume the computational domain under consideration is line segment $[0, D] $, for given $D$. 
We mesh the computational domain by choosing $\Delta x$ to partition interval $[0, D]$.
Let $N_x = D/\Delta x$.
The discrete mesh points $x_{p}$ are given by
\begin{align}
x_p =p \Delta x, \quad x_{p+1/2}=x_{p}+ \frac{\Delta x}{2}=\Big(p+\frac{1}{2}\Big)\Delta x, \label{eq:x_p}
\end{align}
for $p= 0, 1, \dots, N_x$.

In order to simplify notation, let us set $\phi = f^i$, $\theta = \theta_i$, $t_0 = t^k$, $t_f = t^{k+1}$.
Let $M$ be a positive integer ($\geq 3$, in practice). We associate with $M$ a time discretization step $\tau = (t_f - t_0)/M
=\Delta t /M$
and set $t^m = t_0 + m \tau$, with $m = 1, \dots, M$. Next, we present the space and time discretization for each subproblem in Sec.~\ref{lie_scheme}.
\bigskip

\noindent{\bf Step 1}

\noindent With the simplified notation, problem \eqref{eq:1dstep1} can be rewritten as
\begin{equation}
\begin{cases}
 \dfrac{\partial \phi}{\partial t} + \dfrac{\partial }{\partial x} \left( (v [\rho] \cos\theta) \phi(t, x) \right)=0
   \,\,\, \text{on } (t_0, t_f), \\ \label{eq:step1_bis}
\phi(t_0, x)= \phi_0.
\end{cases} 
\end{equation}
For the space discretization, we use a finite difference method which produces an approximation $\Phi_{p}^{m} \in \mathbb{R}$ 
of the cell average 
\[
\Phi_{p}^{m} \approx \dfrac{1}{\Delta x} \int_{x_{p-1/2}}^{x_{p+1/2}}
\phi(t^m, x) dx, 
\]
where $m=1, \dots, M$, $1 \leq p \leq N_x-1$. 
Given an initial condition $\phi_0$,  function $\phi^m$ will be approximated by $\Phi^{m}$ with
\[
\Phi^m \bigg|_{[x_{p-1/2}, \, x_{p+1/2}]} = \Phi_{p}^{m}
\]

The Lax-Friedrichs method for problem \eqref{eq:step1_bis} can be written in conservative form as follows:
\[\Phi_{p}^{m+1}=\Phi_{p}^{m}- \dfrac{\tau}{\Delta x}\Big(  \mathcal{F}(\Phi_{p}^{m}, \Phi_{p+1}^{m}) 
- \mathcal{F}(\Phi_{p-1}^{m}, \Phi_{p}^{m}) \Big)\]
where
\[ \mathcal{F}(\Phi_{p}^{m}, \Phi_{p+1}^{m}) =\dfrac{\Delta x}{2\tau}(\Phi_{p}^{m}- \Phi_{p+1}^{m}) + \dfrac{1}{2} \Big( (v [\rho^m_{p}] \cos\theta) \Phi_{p}^{m}+(v [\rho_{p+1}^m] \cos\theta) \Phi_{p+1}^{m} \Big).  \]
\bigskip

\noindent{\bf Step 2}

\noindent Let $\mathcal{J} = \mathcal{J}^i $ and $\phi_0 = f^{i,k+\frac{1}{2}}$.
Problem \eqref{eq:1dstep3} can be rewritten as
\[
\begin{cases}
 \dfrac{\partial \phi}{\partial t} = \mathcal{J}[f](t, x)  \,\,\, \text{on } (t_0, t_{f}), \cl
\phi(t_0, x)= \phi_0. \el
\end{cases}
\]
For the approximation of the above problem, we use the Forward Euler scheme:
\[ \Phi_{p}^{m+1}= \Phi_{p}^{m} + \tau \Big (\mathcal{J}^m[F^{m}] \Big ). \] 
\bigskip

\subsection{Numerical results} \label{chap:1dnumericalresults}

\begin{figure}[h]
\centering
\begin{overpic}[width=0.49\textwidth,grid=false,tics=10]{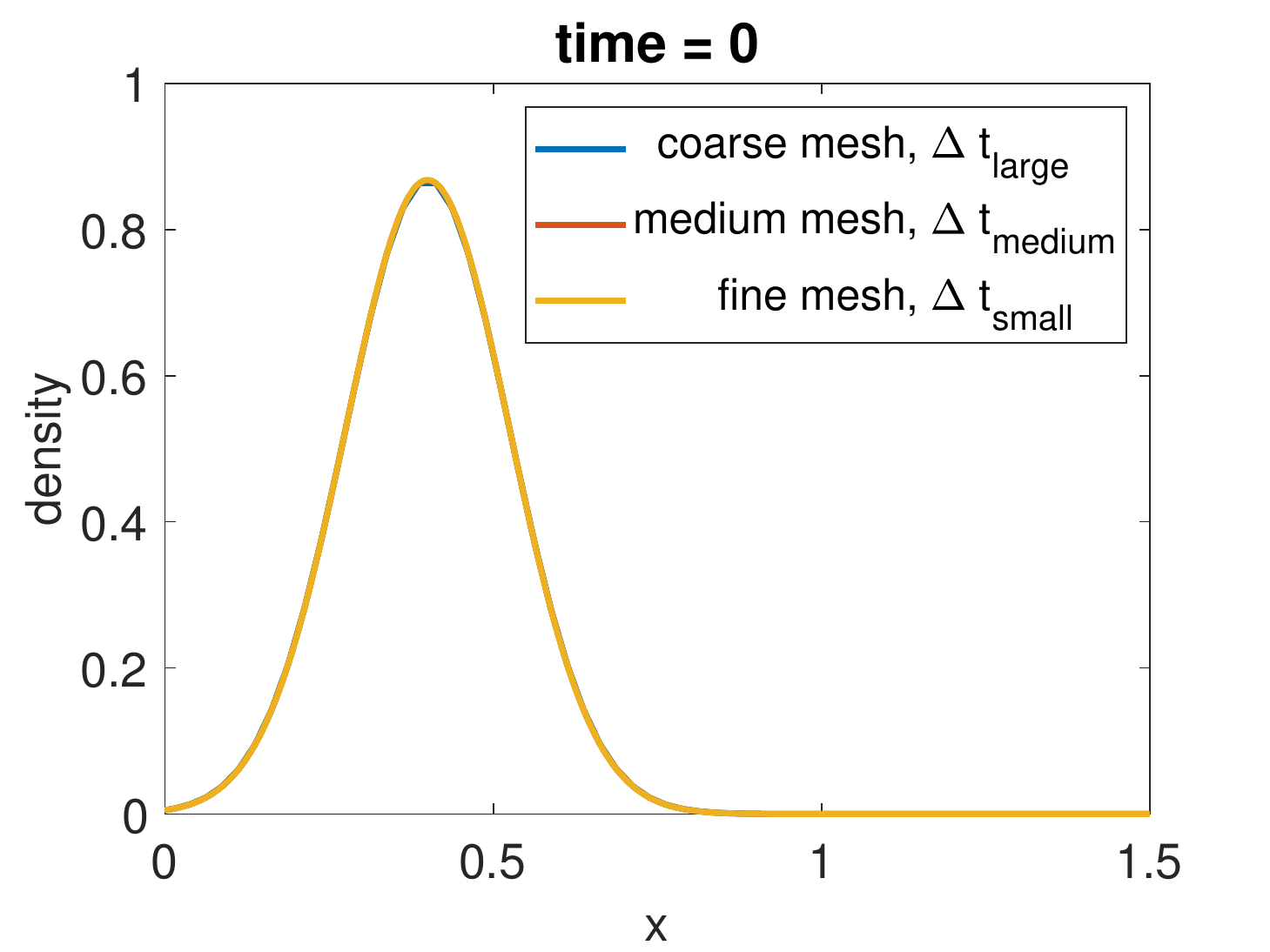}
\linethickness{1pt}
\put(64.5, 22){\vector(0,-1){10}}
\put(61, 25){\text{exit}}
\end{overpic} 
\begin{overpic}[width=0.49\textwidth,grid=false,tics=10]{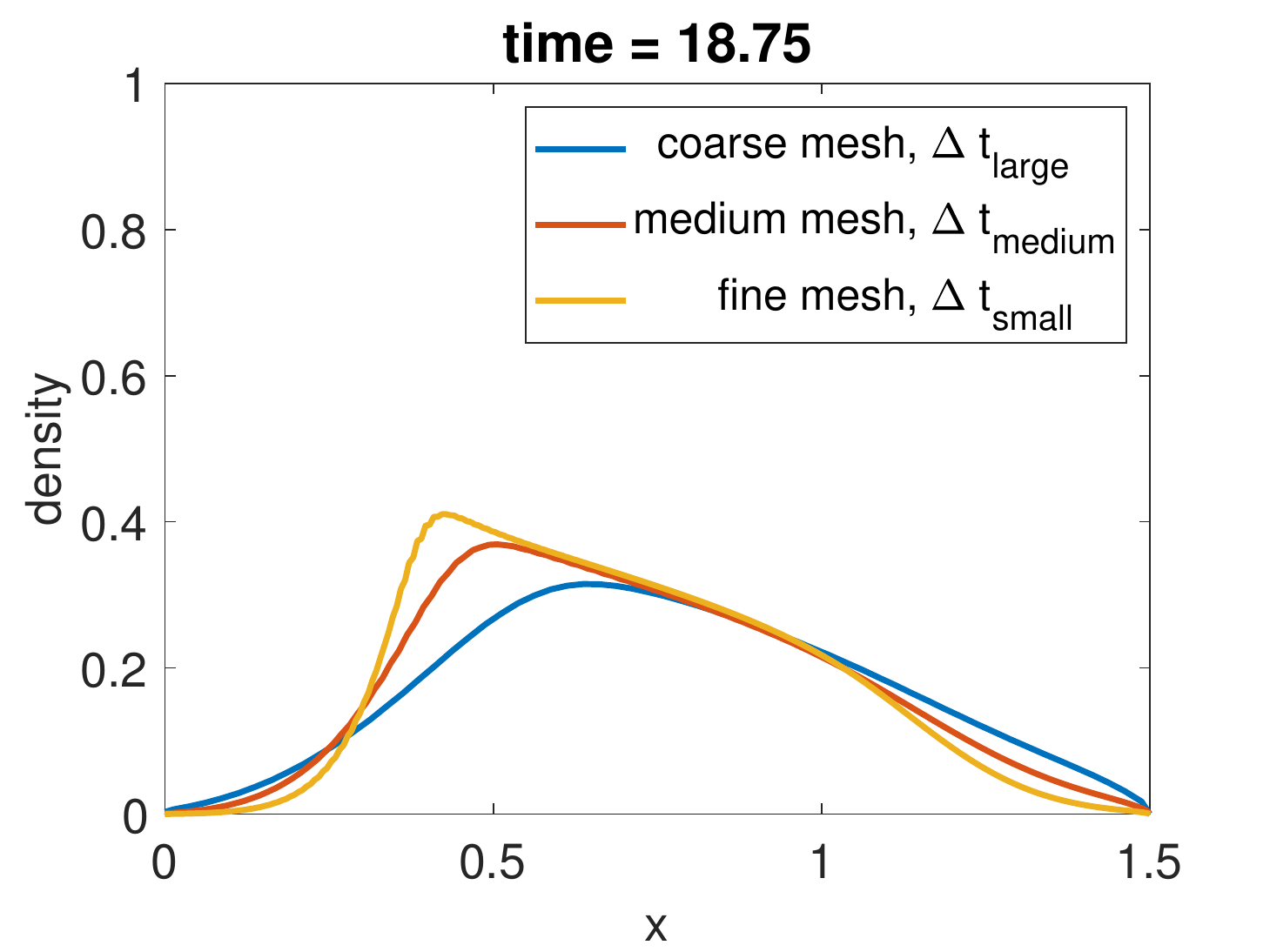}
\end{overpic}
\begin{overpic}[width=0.49\textwidth,grid=false,tics=10]{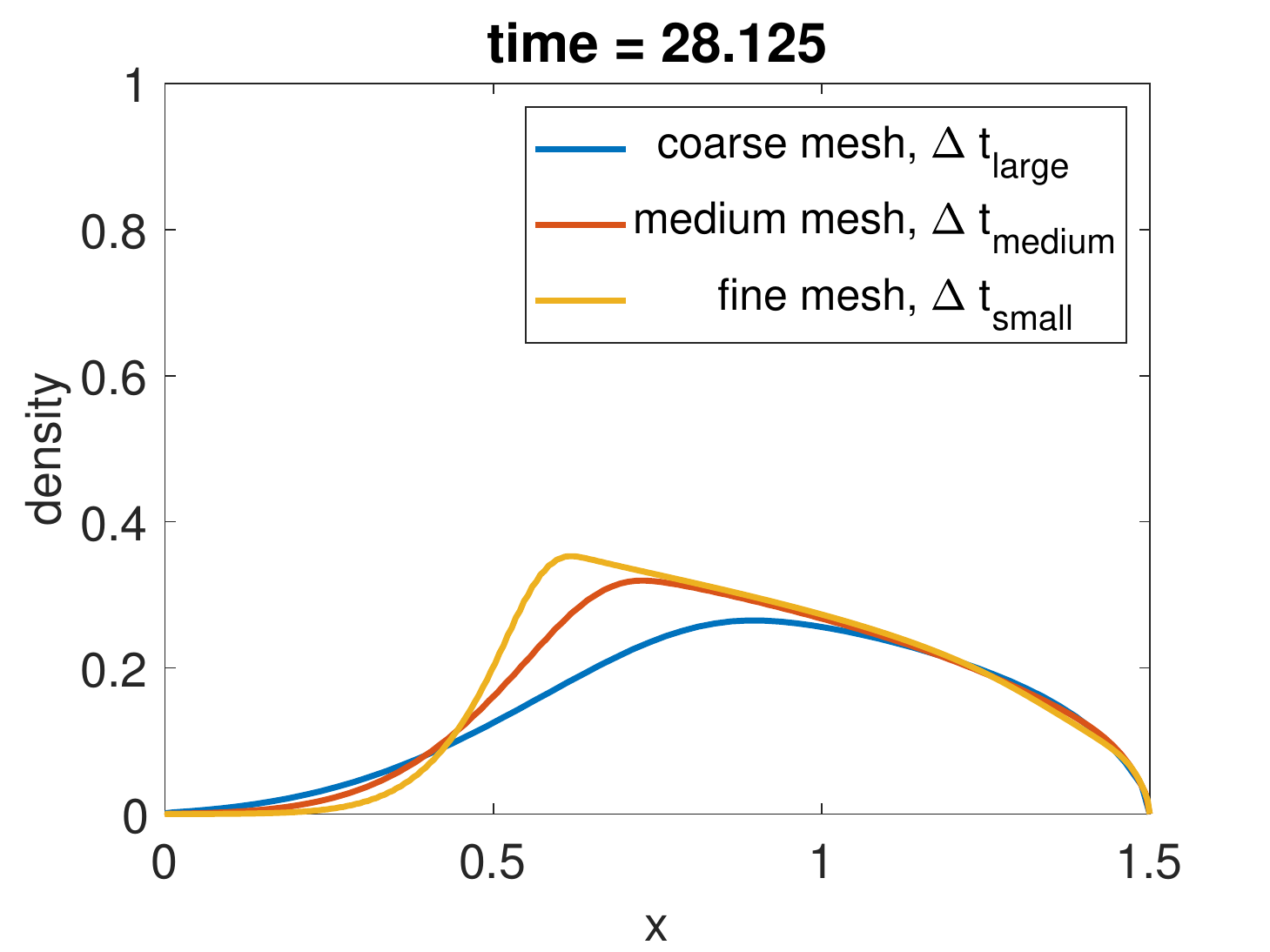}
\end{overpic}
\begin{overpic}[width=0.49\textwidth,grid=false,tics=10]{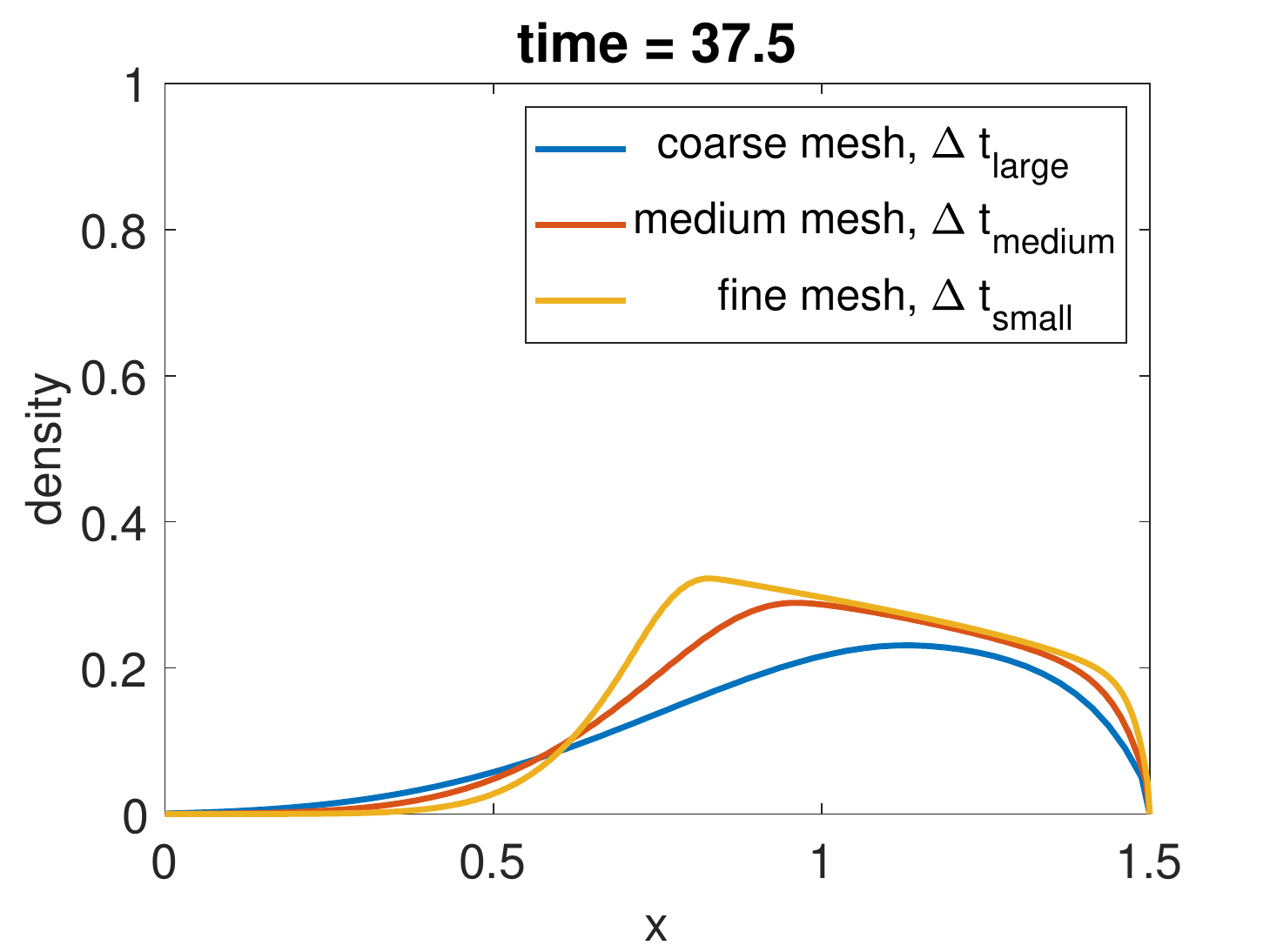}
\end{overpic}
\caption{Evacuation process of 34 pedestrians with initial direction $\theta_1$: density 
for $t= 0, 18.75, 28.125,$ and $37.5$ s computed with three different combinations of mesh and time step.} 
\label{CL_1d}
\end{figure}

We test the algorithm presented in Sec.~\ref{sec:1d-discretization} on a
a simple problem. The computational domain is [0, 150] m, with an exit placed at $x_{exit} = 100$ m. 
A group of $34$ pedestrians is initially located as shown in Fig.~\ref{CL_1d} top, left panel.
Pedestrians are moving toward the exit with the initial direction $\theta_1$. 
This simulation is performed with $\varepsilon=0.4$ and $\alpha=1$. 
We define the following reference quantities: $D=100$ m, 
$v_M= 2$ m/s, $T = D/v_M= 50$ s,  $\rho_M = 3$ people/$m$.

We use three different meshes: a coarse mesh with $\Delta x =1$ m,
a medium mesh with $\Delta x =0.5$ m, and a fine mesh with $\Delta x =0.25$ m.
Similarly, for the time discretization we consider three different time steps:
a large time step $\Delta t_{large} = 0.03$ s,  a medium time step $\Delta t_{medium} = 0.015$ s, and
a small time step $\Delta t_{small} = 0.0075$ s.
The value of $M$ is set to 3.
To avoid stability issues, we consider the following three combinations of the above meshes and time steps:
\begin{enumerate}
\item coarse mesh and $\Delta t_{large}$; 
\item medium mesh and $\Delta t_{medium}$;
\item  fine mesh and $\Delta t_{small}$.
\end{enumerate}
Fig.~\ref{CL_1d} shows the computed density at four different times for the three combinations mentioned above.  
We see that the density computed for medium mesh-$\Delta t_{medium}$ is close to the density computed 
with fine mesh-$\Delta t_{small}$ almost on the entire computational domain, indicating convergence. 

\section{Contagion model}\label{sec:emotional}

In this section, we briefly recall the ASCRIBE model as presented in \cite{Bertozzi2015} and validate our implementation of it. 
The contagion modeled by ASCRIBE is emotional: it assumes that the pedestrian's velocity is equal
to the fear level \cite{Bertozzi2015}, i.e~they walk faster if scared.
For a mathematical analysis of the ASCRIBE model through particle, continuum, and kinetic descriptions we
refer to \cite{Bertozzi2014ContagionSI}. See also \cite{Bellomo2013_new, Bellomo2015_new}.
We take inspiration from ASCRIBE to model disease contagion.

We consider a group of $N$ pedestrians. 
Let $x_m(t)$ and $v_m(t)$ denote the position and velocity of pedestrian $m$.
We denote with $q_m$ the pedestrian's contagion level, with the meaning of fear contagion in ASCRIBE and 
disease contagion in our model.
The microscopic description of the pedestrian motion is given by: 
{\small
\begin{equation}\label{eq:fearf}
\frac{dx_m}{dt}=v_m; \quad \frac{dq_m}{dt}=\gamma(q_m^*-q_m); \quad q_m^* =\frac{\sum_{j=1}^{N} \kappa_{m,j}q_j}{\sum_{j=1}^{N}\kappa_{m,j}}, \quad  m= 1,2,3, \dots, {N}.
\end{equation}
} 
The quantity $q_m^{\ast}$ is a weighted local ``average'' contagion level. 
In \eqref{eq:fearf}, $\kappa_{m,j}=\kappa(|x_m-x_j|)$ is an interaction kernel, which serves as the weight 
in the average $q_m^{\ast}$. 
The interaction kernel $\kappa$ is a decreasing function of the mutual distance between two particles and is parametrized by an interaction distance $R$, for example:
\begin{equation}\label{eq:kernel}
\kappa(x) = \cfrac{R}{(x^2+R^2)\pi}.
\end{equation}
Parameter $\gamma \geq 0$ in eq.~\eqref{eq:fearf} describes the contagion interaction strength and its value could be particle-dependent
for more general cases. The model works as follows: for $\gamma = 0$ there is no contagion, while for
$\gamma \neq 0$ the contagion is faster the larger the value of $\gamma$. 

Following \cite{Bertozzi2015}, we recall the empirical distribution density by
\begin{equation}
h^{N}(t,x,q)=\frac{1}{N}\sum_{m=1}^{N}\delta(x-x_m(t))\delta(q-q_m(t)), \cl
\end{equation}
where $\delta$ is the Dirac delta function.
We assume that the pedestrians remain in a fixed compact domain $(x_m(t), q_m(t)) \in \Omega \subset \mathbb{R}^2$ for all $m$ and up to the time we consider. Let $\langle \cdot \rangle_{x,q}$ denote integration against both $x$ and $q$. 
Considering a test function $\psi \in C^1_0(\Omega)$, we have:
\begin{align} \label{eq:testfuction}
\frac{d}{dt} \langle h^{N}, \psi \rangle_{x,q} 
&=\frac{d}{dt} \biggl< \frac{1}{N}\sum_{m=1}^{N}\delta(x-x_m(t))\delta(q-q_m(t)), \psi \biggr>_{x,q} =\frac{d}{dt} \frac{1}{N}\sum_{m=1}^{N}\psi(x_m(t), q_m(t)) \cl
&=\frac{1}{N}\sum_{m=1}^{N} \left( \psi_x v_m+ \psi_q\gamma(q_m^{\ast}-q_m) \right) = \langle \psi_xv_m, h^{N} \rangle + \frac{\gamma}{N}\sum_{m=1}^{N}\psi_q \biggl( \frac{\sum_{j=1}^{N} \kappa_{m,j}q_j}{\sum_{j=1}^{N}\kappa_{m,j}}- q_m \biggr).
\end{align} 
As shown in \cite{Bertozzi2015}, eq.~\eqref{eq:testfuction} leads to the limiting kinetic equation
\begin{equation} \label{eq:disease_kineticeq}
h_t+(vh)_x=\gamma((q-q^\ast)h)_q,
\end{equation}
where $q^\ast$ is a weighted average contagion level.
Like in the case for the pedestrian dynamics model, we will make use of dimensionless quantities. For this purpose, 
in addition to the reference quantities in Sec.~\ref{sec:ped_model} we denote with $q_{max}$ the highest fear level.

In this section, we will consider two nondimensional velocity moduli for eq.~\eqref{eq:disease_kineticeq}:
\begin{equation}\label{eq:v_q}
v = q \quad \text{and} \quad v = (1-q)^2.
\end{equation}
The first modulus corresponds to the ASCRIBE model: people walk faster if scared. With the second
modulus, we model the spreading of a fictitious disease that affects people's walking ability upon contagion:
people slow down when $q$ increases.


\subsection{Full discretization}\label{sec:contagion_full_disc}
We present a space and time discretization for eq.~(\ref{eq:disease_kineticeq}).
We partition the spatial domain into subdomains $[x_{j-\frac{1}{2}}, x_{j+\frac{1}{2}}]$,
with $j \in 1,2,\dots, N_x$, of equal length $\Delta x$. See \eqref{eq:x_p}. 
The contagion level domain is partitioned into subdomains $[q_{l-\frac{1}{2}}, q_{l+\frac{1}{2}}]$, with $l \in 1,2,\dots, N_q$, where 
\begin{align}
q_l =l \Delta q, \quad q_{l+1/2}=q_{l}+ \frac{\Delta q}{2}=\Big(l+\frac{1}{2}\Big)\Delta q. \el
\end{align}
For simplicity, we assume that all subintervals have equal length $\Delta q$.
The two partitions induce a partition of domain $\Omega$ into cells.
The time step $\Delta t$ is chose as 
\[ \Delta t = \frac{1}{2}\min\Bigg\{ \frac{\Delta x}{\max_{j}q_{j}}, \frac{\Delta q}{2\gamma \max_{j}q_{j}} \Bigg\}\] 
to satisfy the Courant-Friedrichs-Lewy (CFL) condition.

Let us denote $h_{j, l}=h(t, x_j, q_l)$ and $q_j^ \ast=q^\ast(t, x_j)$.
We consider a first-order semi-discrete upwind scheme for eq.~\eqref{eq:disease_kineticeq} 
adapted from one of the methods used in \cite{Bertozzi2015}. This scheme is
second-order in velocity thanks to the use of a flux limiter and it reads:
\begin{equation} \label{eq:disease_modified}
\partial_t h_{j, l}+\frac{\eta_{j+\frac{1}{2}, l}-\eta_{j-\frac{1}{2}, l}}{\Delta x}
+\gamma \frac{\xi_{j, l+\frac{1}{2}}-\xi_{j, l-\frac{1}{2}}}{\Delta q} 
+\gamma \frac{C_{j, l+\frac{1}{2}}-C_{j, l-\frac{1}{2}}}{\Delta q}= 0,
\end{equation}
where
\begin{align}
\eta_{j, l+\frac{1}{2}} & = v_lh_{j, l},\cl
\xi_{j, l+\frac{1}{2}} & = \dfrac{|q^\ast_j-q_{l+\frac{1}{2}} |+(q^\ast_j-q_{l+\frac{1}{2}})}{2}h_{j, l}+ \dfrac{|q^\ast_j-q_{l+\frac{1}{2}} |-(q^\ast_j-q_{l+\frac{1}{2}})}{2}h_{j, l+1}, \cl
C_{j, l+\frac{1}{2}} &=\frac{1}{2} \left| s_{j, l+\frac{1}{2}} \right| \left(1-\frac{\Delta t}{\Delta q} \left| s_{j, l+\frac{1}{2}} \right| \right) W_{j, l+\frac{1}{2}}\varphi \left( \frac{W_{j,{b}+\frac{1}{2}}}{W_{j, l+\frac{1}{2}}}\right) \label{eq:C}
\end{align}
with 
\begin{equation}
s_{j, l+\frac{1}{2}}= q^{\ast}_j-q_{l+\frac{1}{2}},~\text{and}~W_{j, l+\frac{1}{2}}=h_{j, l}-h_{j, l+\frac{1}{2}}.
\end{equation}
In \eqref{eq:C}, the subscript ${b}$ in \eqref{eq:C} is $l-1$ if $s_{j, l-\frac{1}{2}} > 0$ and $l+1$ if $s_{j, l-\frac{1}{2}} < 0$, and 
$\varphi$ is a slope limiter function. We choose the Van Leer function:
\begin{equation}\label{van_leer}
\varphi(\theta)=\frac{|\theta|+\theta}{1+|\theta|}.
\end{equation}
See, e.g., \cite{leveque1990numerical} for more details.

Finally, we use the forward Euler scheme for the time discretization of problem~\eqref{eq:disease_modified}:
\begin{equation}
 h_{j,l}^{m+1} =  h_{j,l}^{m}-\Delta t \Bigg ( \cfrac{\eta^{m}_{j+\frac{1}{2}, l}-\eta^{m}_{j-\frac{1}{2}, l}}{\Delta x}
+\gamma \cfrac{\xi^{m}_{j, l+\frac{1}{2}}-\xi^{m}_{j, l-\frac{1}{2}}}{\Delta q} 
+\gamma \cfrac{C^{m}_{j, l+\frac{1}{2}}-C^{m}_{j, l-\frac{1}{2}}}{\Delta q} \Bigg ).
\end{equation}

\subsection{Numerical results}\label{sec:num_res_contagion}

We validate our implementation of the scheme presented in the previous section with a test case taken from \cite{Bertozzi2015}.
The computational domain in the $xq$-plane is $[-10, 10] \times [0, 3]$. To discretize this domain, we choose dimensional $\Delta x=\Delta q=0.02$. We consider a relatively small interaction radius and a rather large interaction strength 
(i.e., quick interactions), which are meant to model a dense crowd setting. In particular, we set
$R=0.0002$ and $\gamma=100$. 
The time interval under consideration is $[0, 6.66]$
and the time step $\Delta t$ is set to $0. 001$. 

\begin{figure}[h]
\center
\subfloat[Initial time, $v = q$]{
\begin{overpic}[width=0.33\textwidth,height=0.26\textwidth, grid=false]{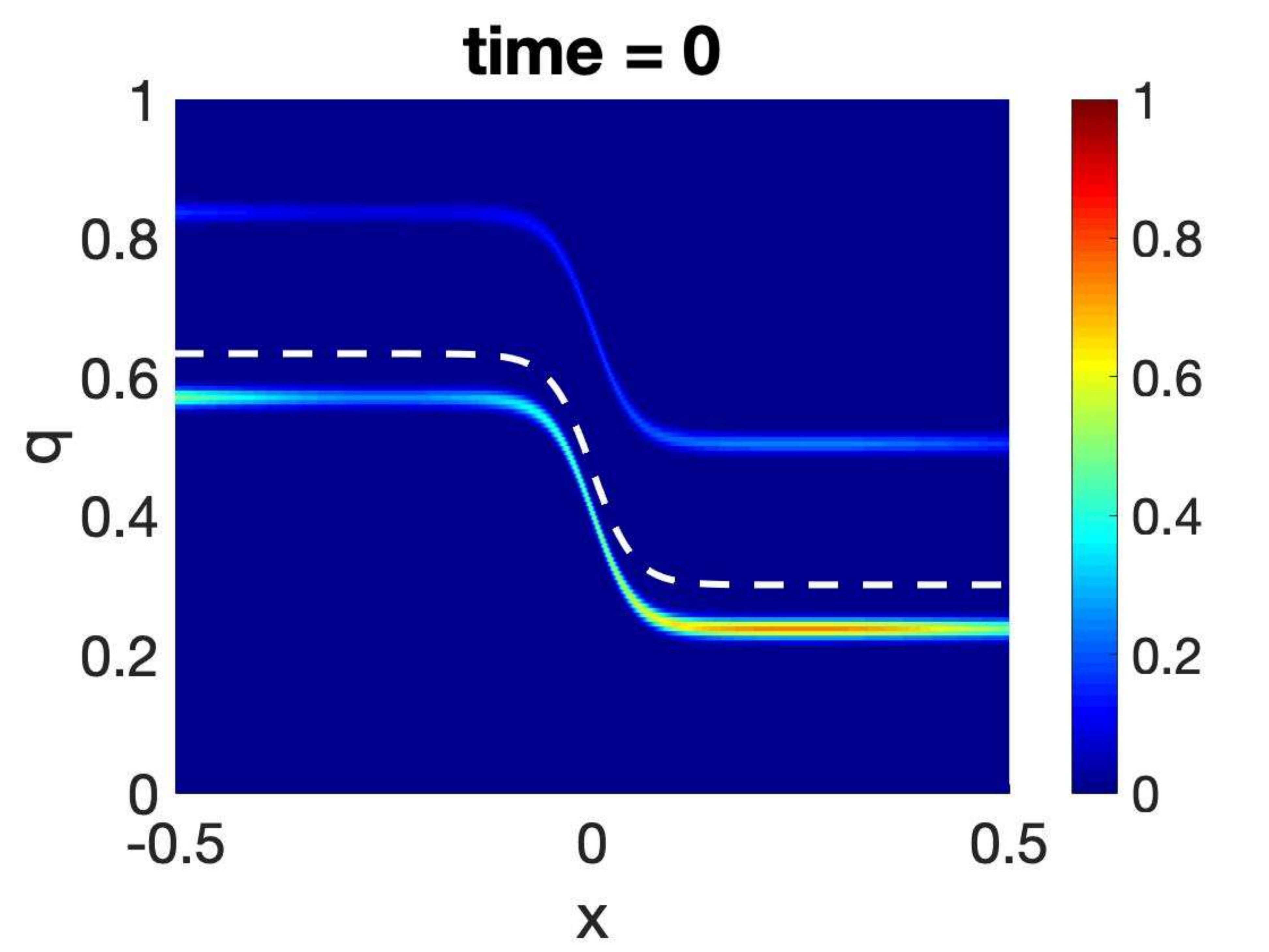}
\end{overpic}
}
\subfloat[Final time, $v = q$]{
\begin{overpic}[width=0.33\textwidth,height=0.26\textwidth,grid=false]{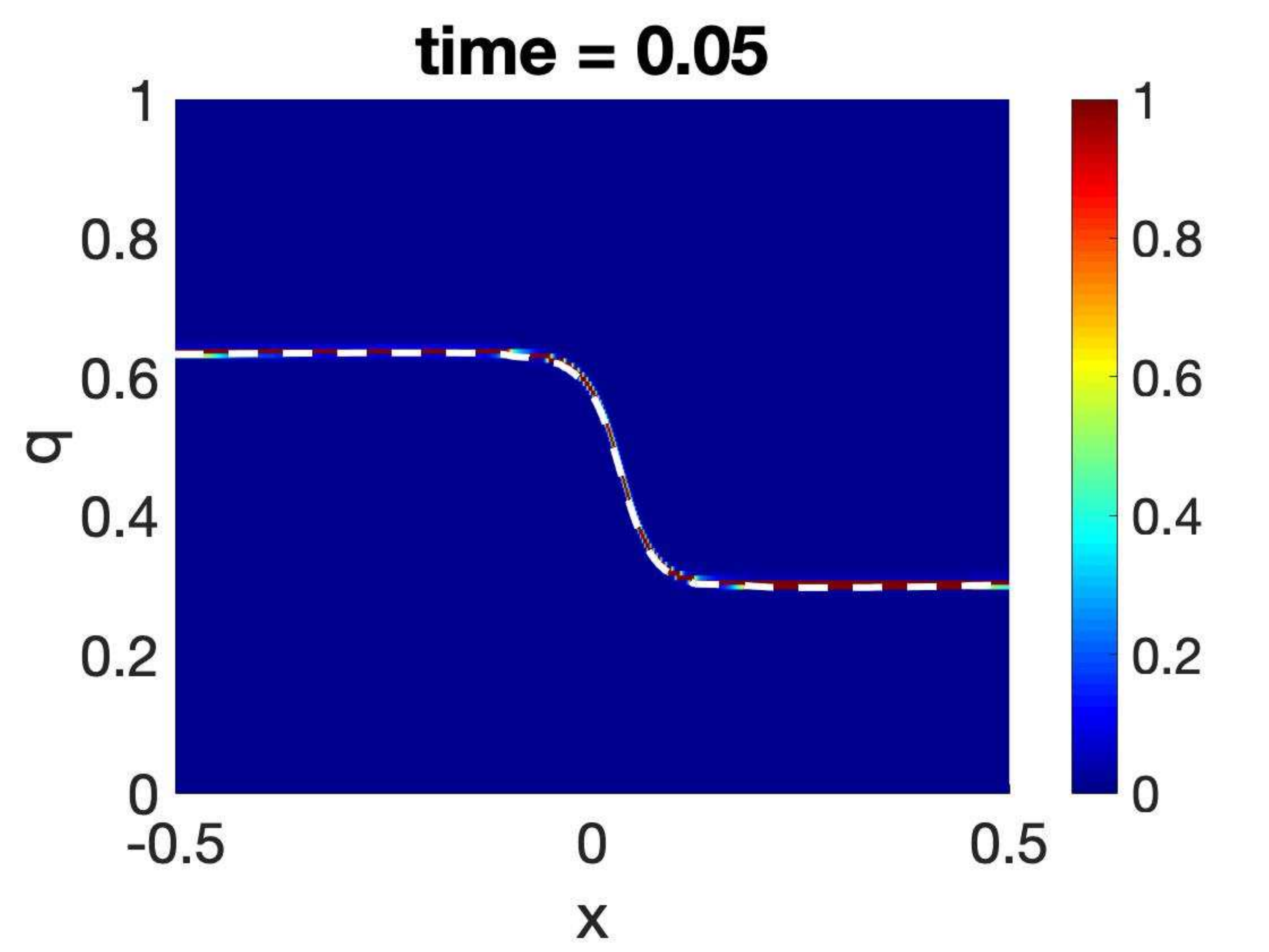}
\end{overpic}
}\\
\subfloat[Initial time, $v = (1 - q)^2$]{
\begin{overpic}[width=0.33\textwidth,height=0.26\textwidth, grid=false,tics=10]{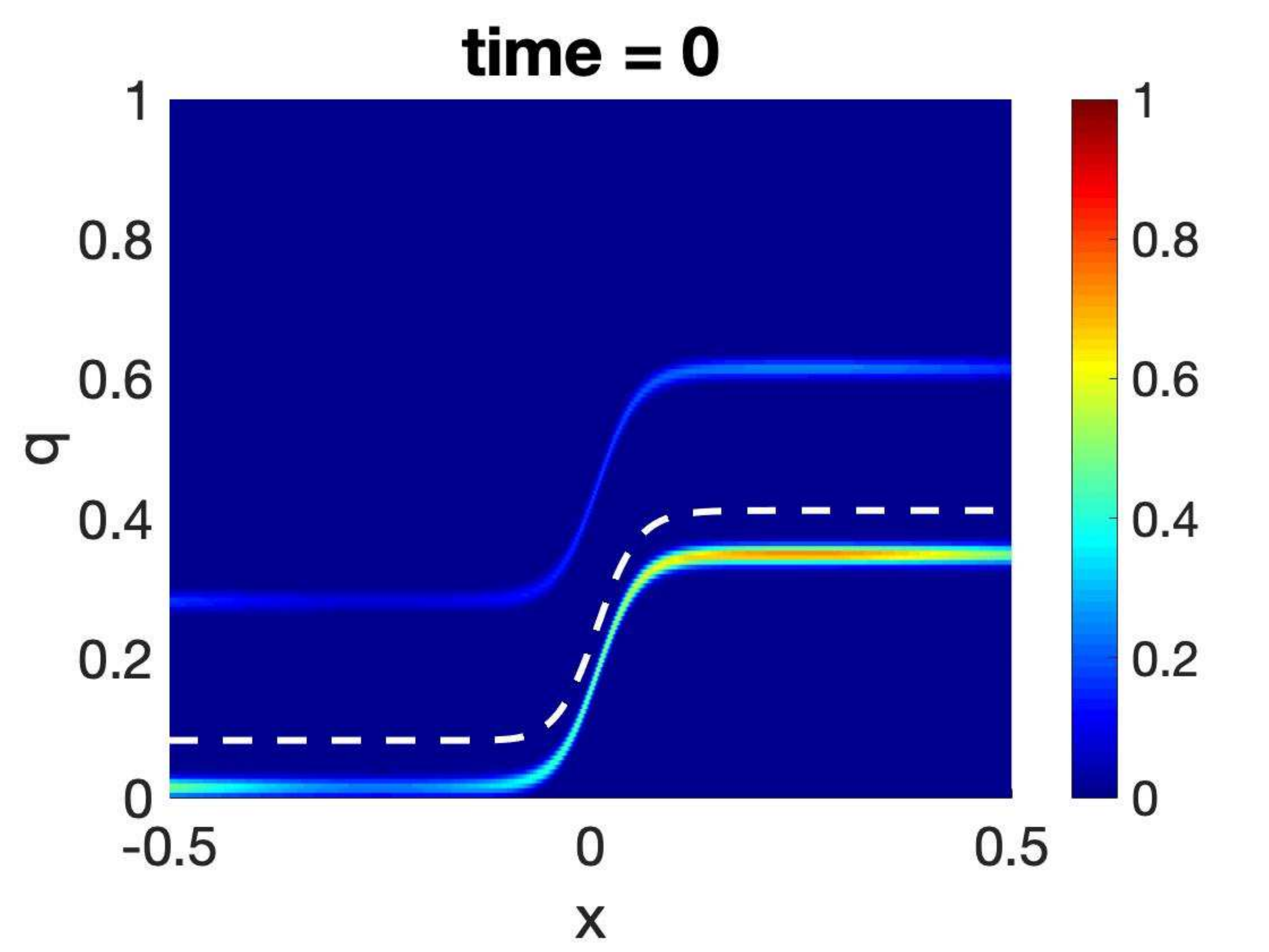}
\end{overpic}
}
\subfloat[Final time, $v = (1 - q)^2$]{
\begin{overpic}[width=0.33\textwidth, height=0.26\textwidth, grid=false,tics=10]{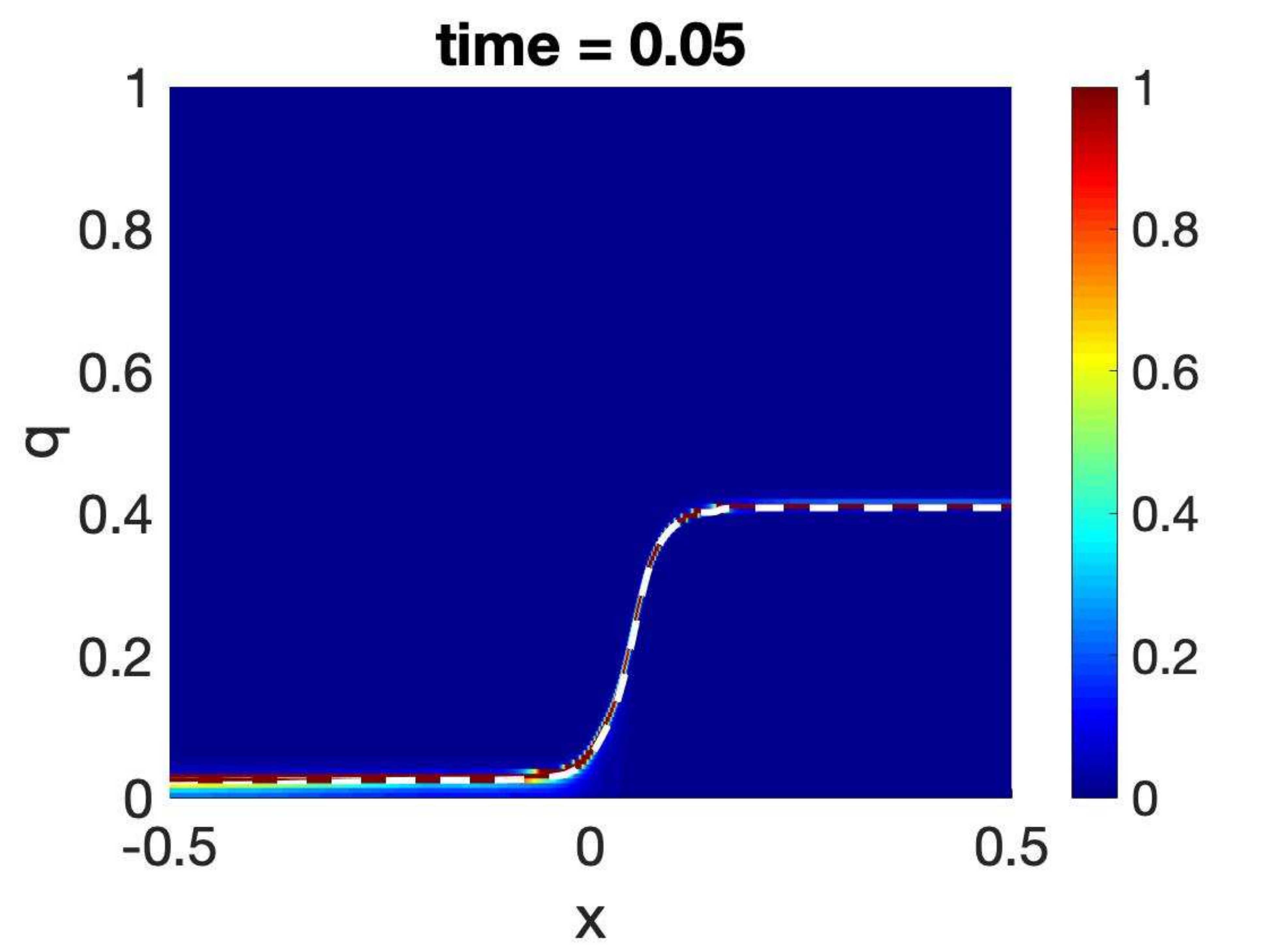}
\end{overpic}
}\\
\caption{Computed distribution density at $t= 0$ s (panels (a) and (c)) and at $t= 0.05$ s (panels (b) and (d)) for
the velocities in \eqref{eq:v_q}. 
The white dash line represents $ q^\ast(t,x)$.
}
\label{Contagion_ref}
\end{figure}

We approximate the delta function as follows:
\begin{equation}
\delta(q) \sim E(q) = \cfrac{1}{\sqrt{\pi}R_0} e^{-\cfrac{q^2}{R_0^2}}, \quad R_0 =0.04.
\end{equation}
Following \cite{Bertozzi2015}, the initial condition for the distribution density is set as: 
\begin{equation}
h(0, x, q) = h_I(x,q)= \left( \sin\left(\cfrac{\pi x}{10} \right)+2 \right)\left( \cfrac{1}{4}\,\,E(q-v_I(x)-0.5)+\cfrac{3}{4}\,\,E(q-v_I(x)+0.3) \right),
\nonumber 
\end{equation}
with
\begin{equation}
q_I(0,x)=\cfrac{1}{2}\Big(3-\tanh x \Big).
\nonumber 
\end{equation}
Fig.~\ref{Contagion_ref} (a) and (b) (resp., Fig.~\ref{Contagion_ref} (c) and (d)) show 
the evolution of $h$ with the first (resp., second) velocity in \eqref{eq:v_q}.
Notice that for both cases, $h_I(x, q)$ has two bumps in $q$ for every $x$.
As time passes, $h(t,x,q)$ starts to concentrate on $q^\ast(t,x)$, as seen in Figure~\ref{Contagion_ref}
(b) and (d). 
Our results in Figure~\ref{Contagion_ref} (a) and (b) are in excellent agreement with the results reported in \cite{Bertozzi2015}.

Let $\Delta Q$ be the greatest difference in contagion level between two people at the initial time.
For the test considered in this section, we have $\Delta Q =1$, and thus $\Delta Q > 2 \gamma R$.
This means that people's paths could cross in the case of $v = q$, as stated in
the following theorem from \cite{Bertozzi2015}:

\noindent {\bf Theorem 1} {\it 
Let us assume that as $R \rightarrow 0$ and $\gamma \rightarrow \infty$, quantity  
$R\gamma$ remains fixed. Let $p_1$ and $p_2$ be two particles located at positions $x_1(0)$ and $x_1(0)+d(0)$, where the initial distance $d(0)\leq R$, 
with velocities $q_1(0)$ and $q_2(0)$, respectively. Then their paths will cross if and only if $q_1(0)-q_2(0) > \gamma d(0)$. Furthermore if $q_1(0)-q_2(0) > \gamma [d(0)+R] $, their paths will cross and they will eventually cease to interact with each other. }

\begin{figure}[h]
\begin{overpic}[width=0.32\textwidth,grid=false,tics=10]{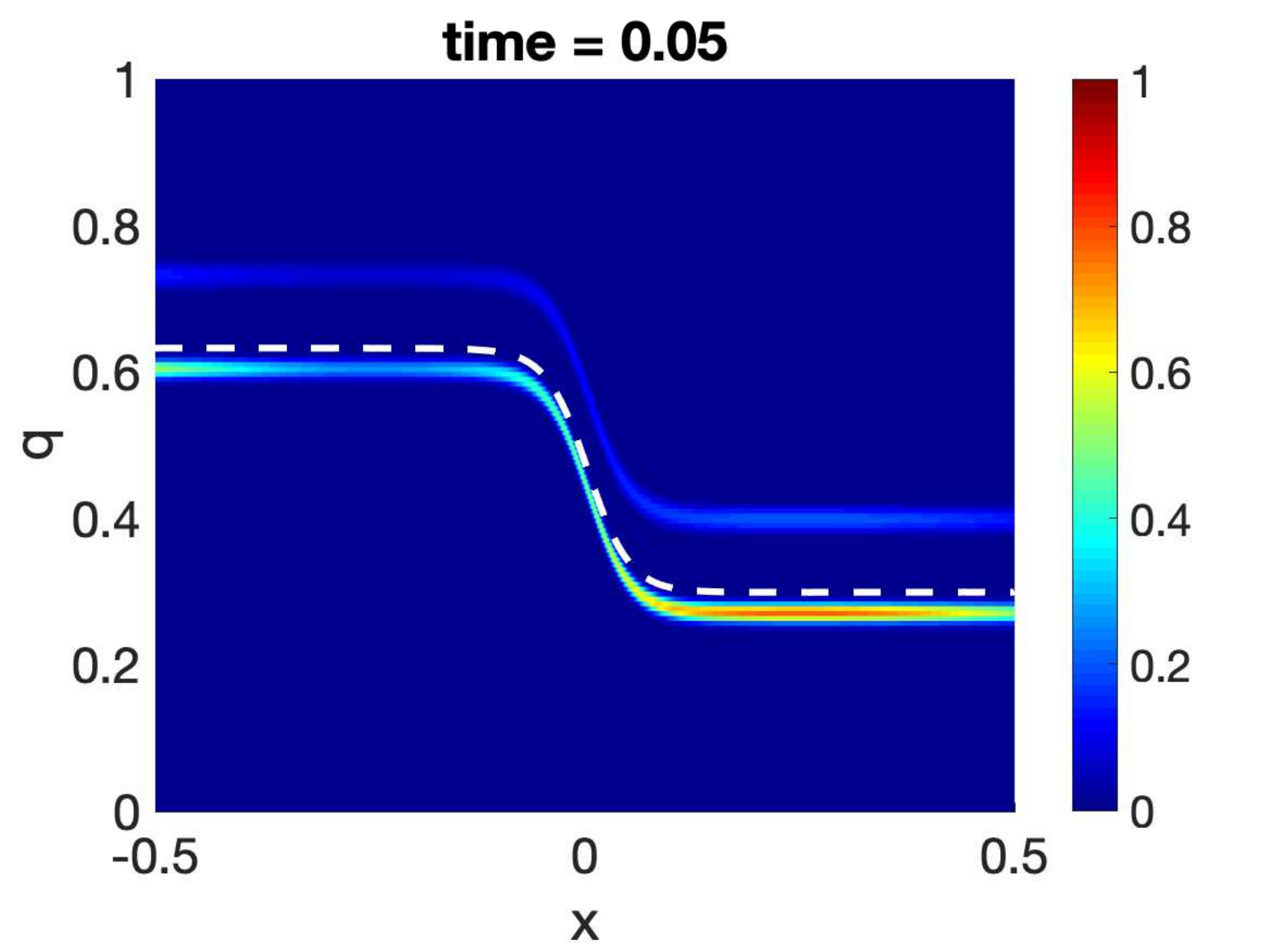}
\end{overpic} 
\begin{overpic}[width=0.32\textwidth,grid=false,tics=10]{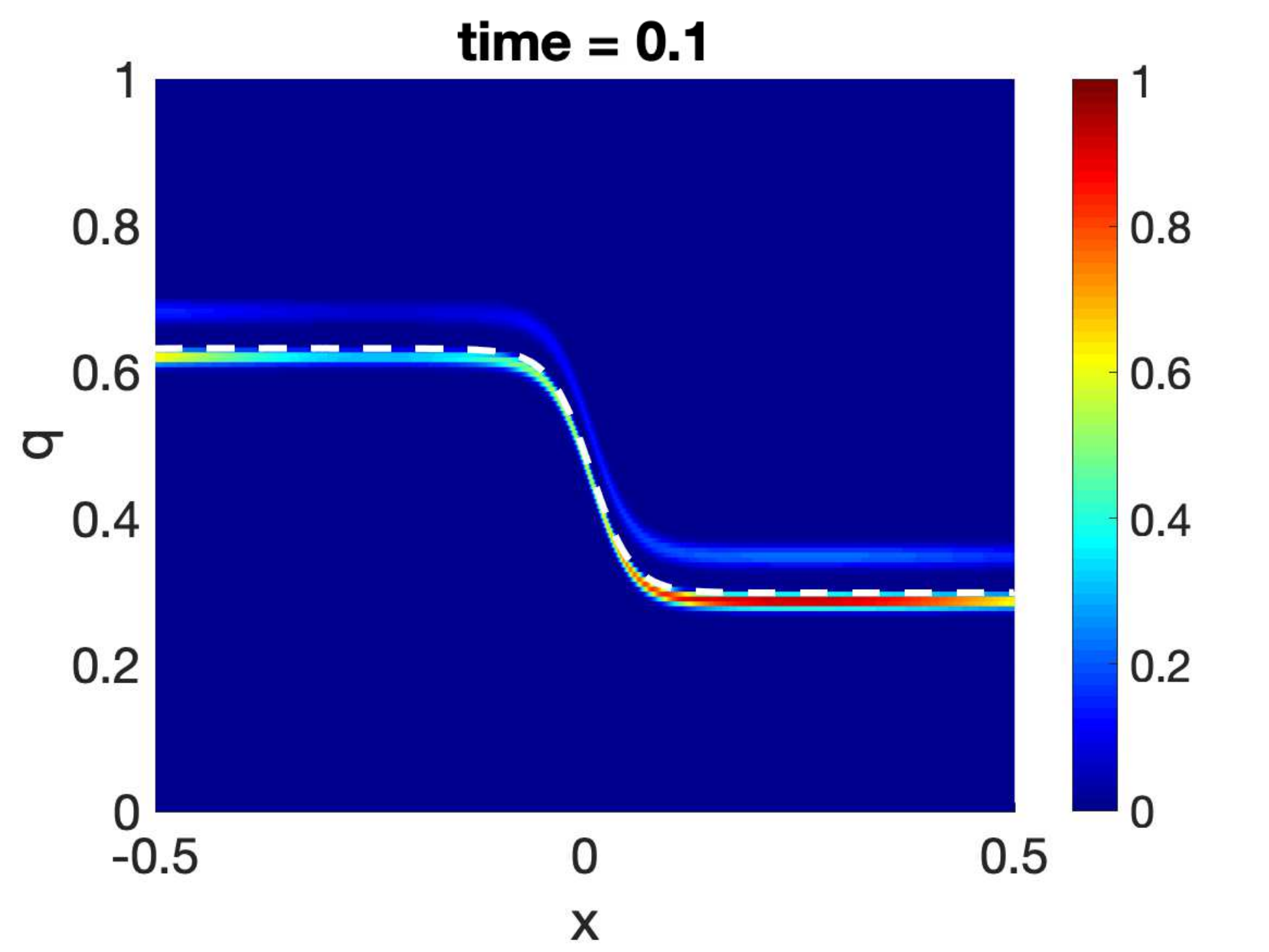}
\put(0,75){$\gamma=100$, $R=0.01$ ($\Delta Q =2 \gamma R$)}
\end{overpic} 
\begin{overpic}[width=0.32\textwidth,grid=false,tics=10]{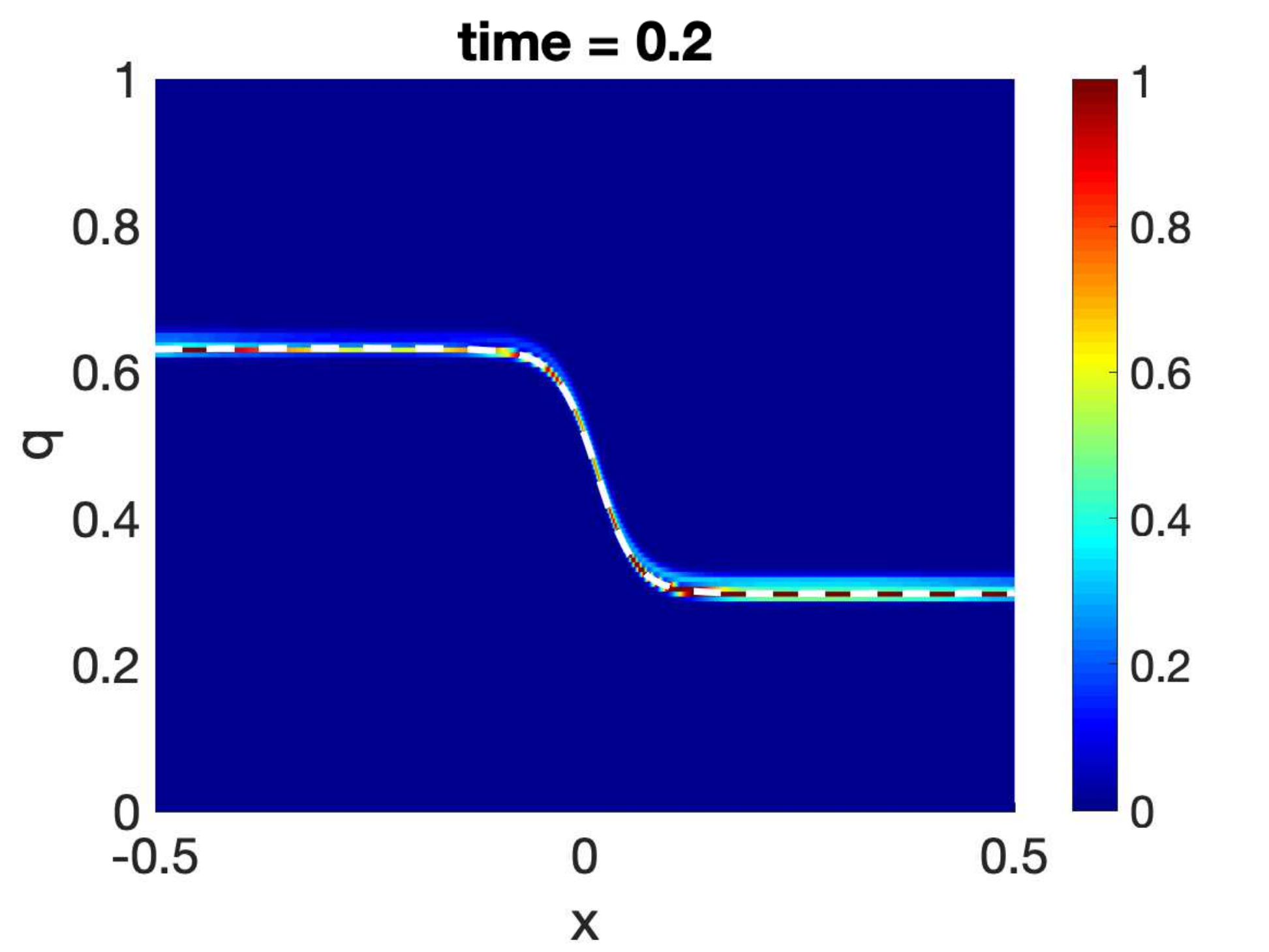}
\end{overpic} 
\vskip .5cm
\begin{overpic}[width=0.32\textwidth,grid=false,tics=10]{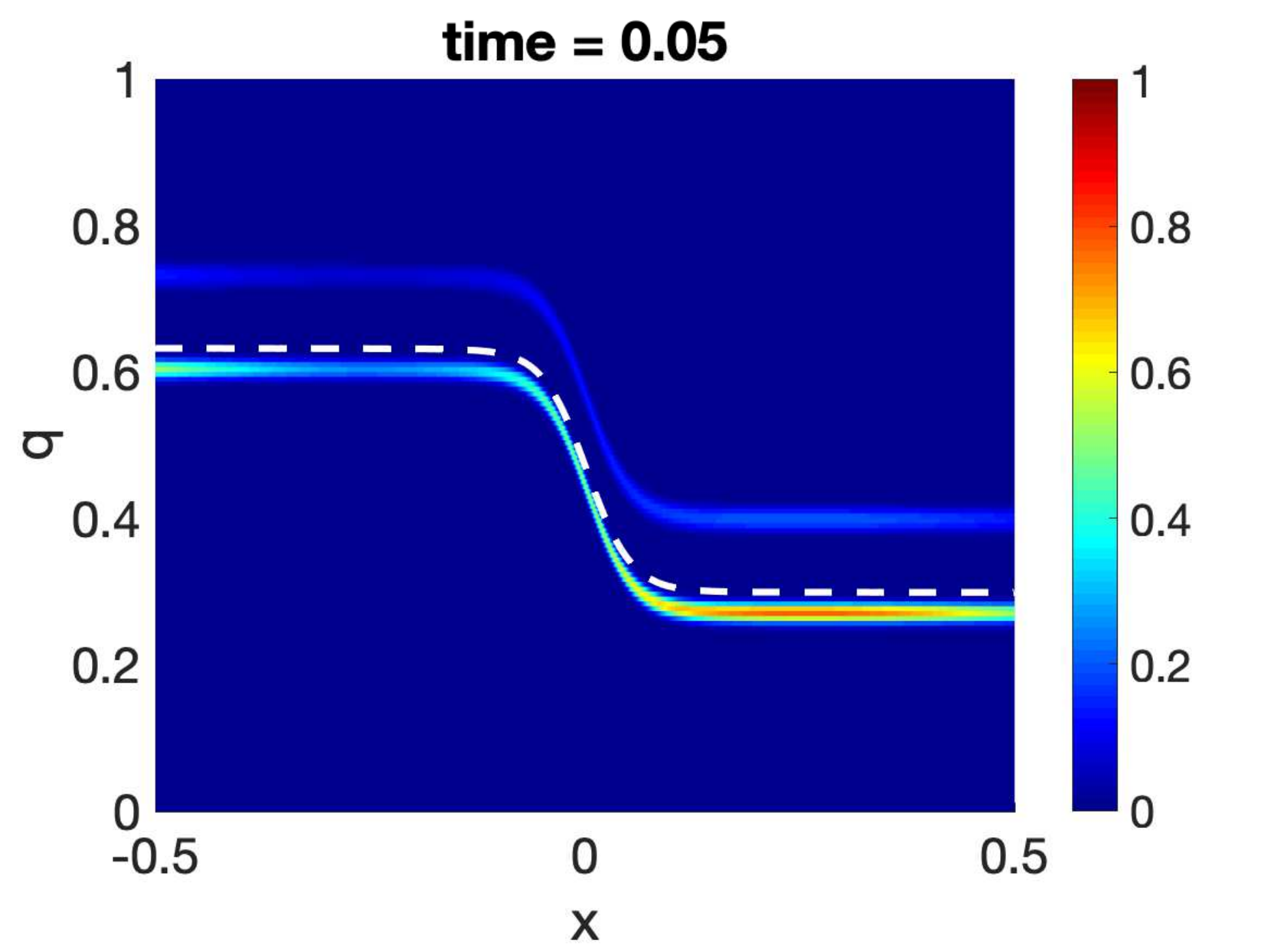}
\end{overpic} 
\begin{overpic}[width=0.32\textwidth,grid=false,tics=10]{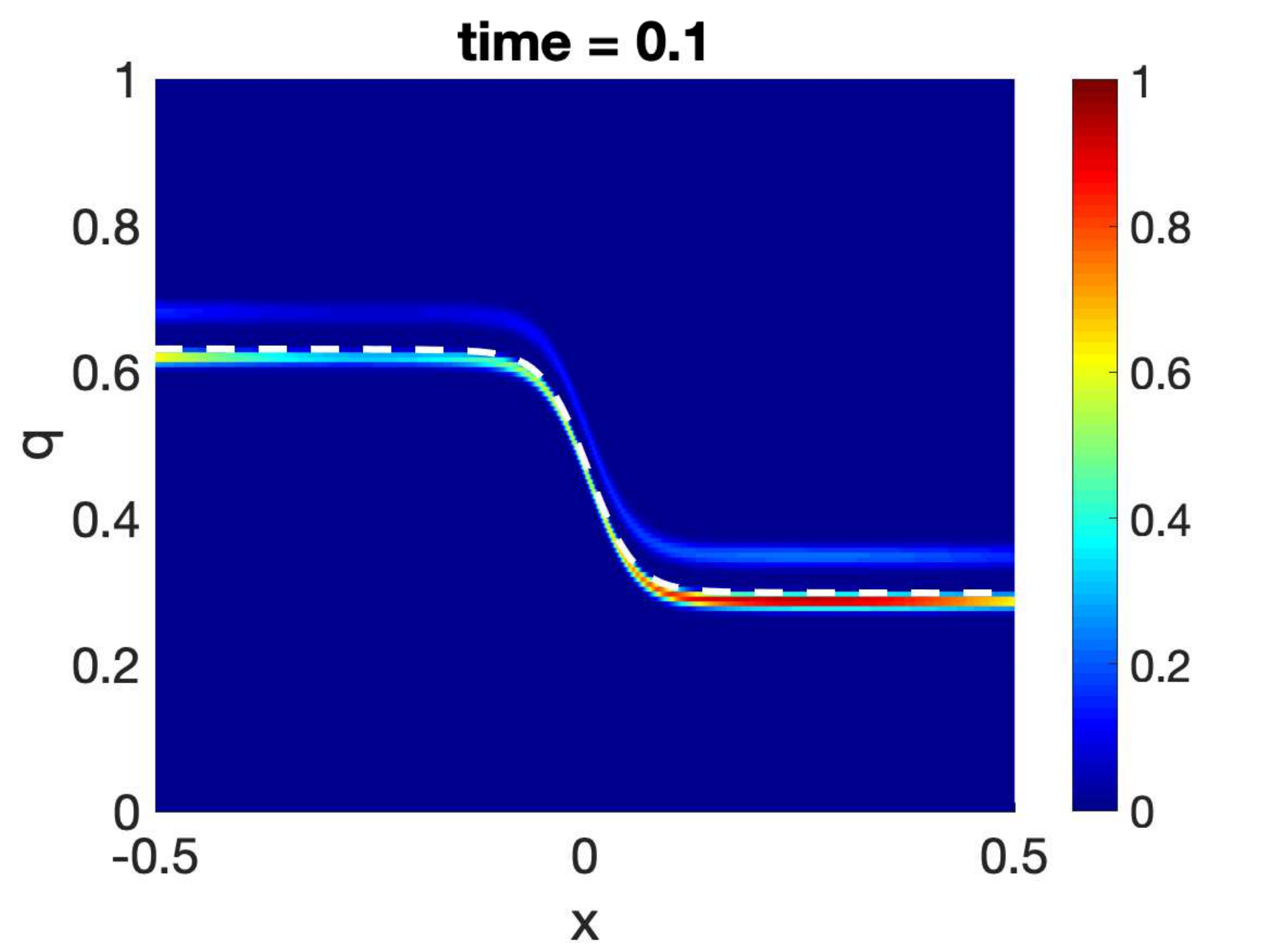}
\put(0,76){$\gamma=100$, $R=0.02$ ($\Delta Q < 2 \gamma R$)}
\end{overpic} 
\begin{overpic}[width=0.32\textwidth,grid=false,tics=10]{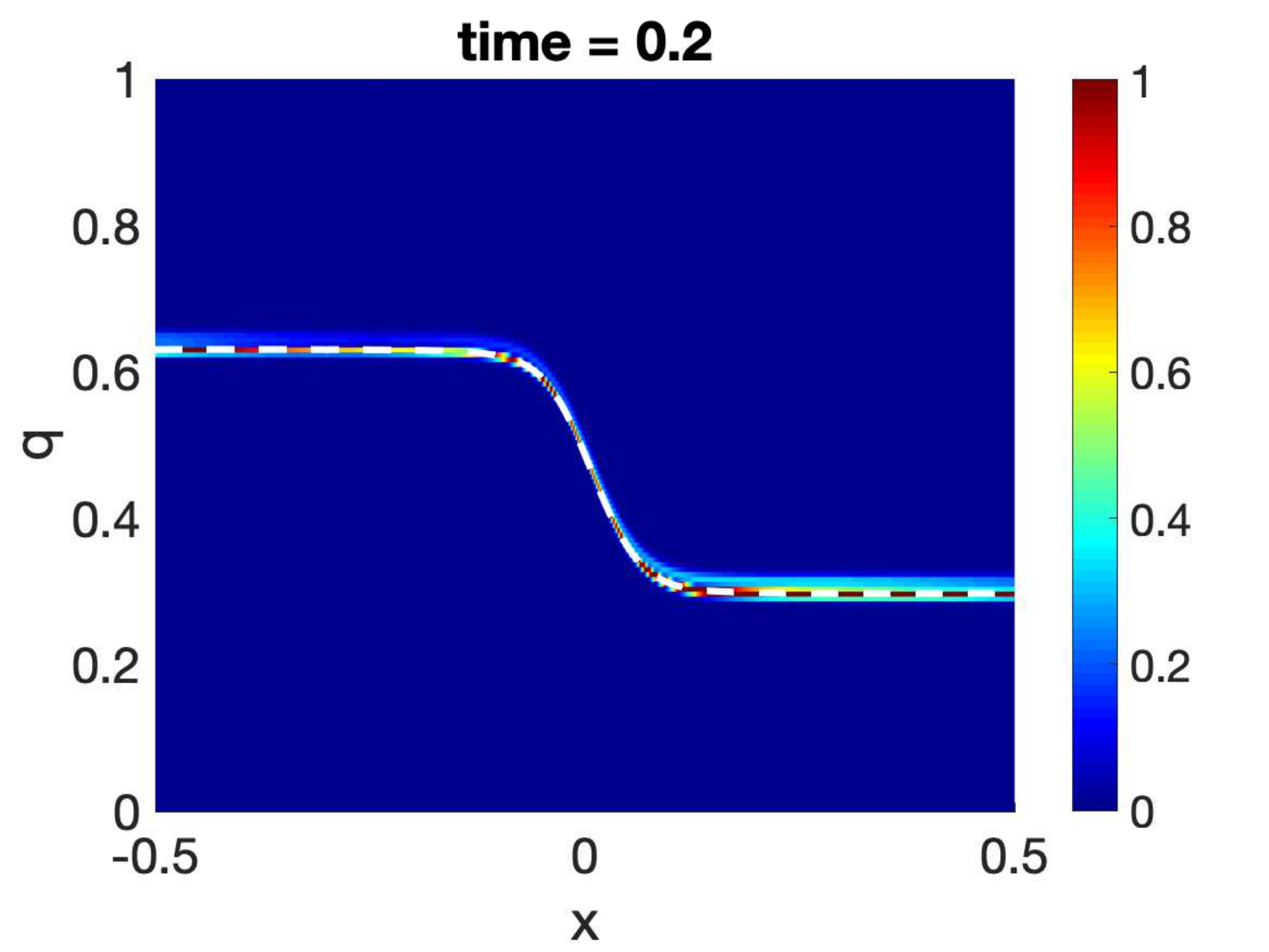}
\end{overpic} 
\vskip .5cm
\begin{overpic}[width=0.32\textwidth,grid=false,tics=10]{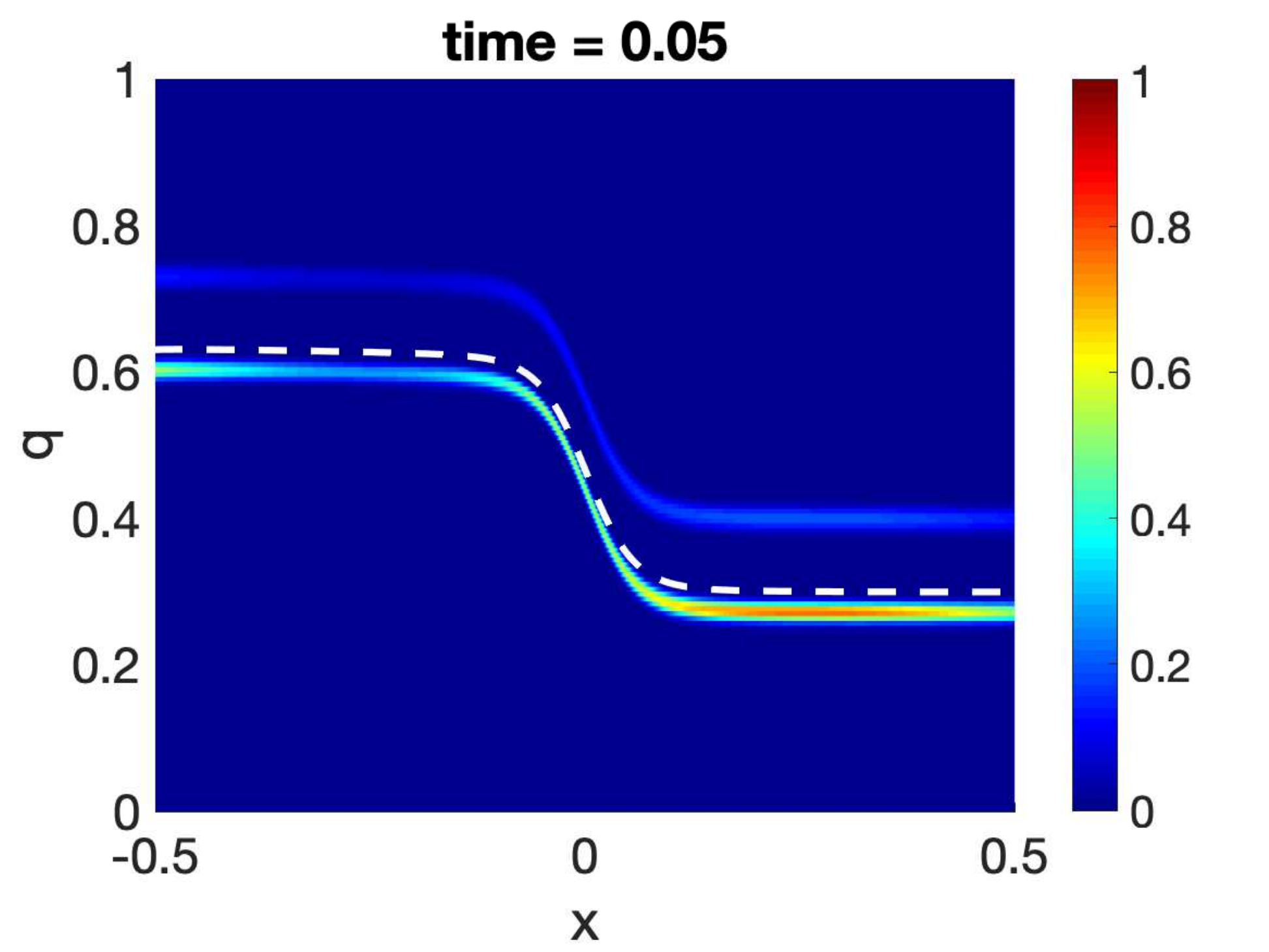}
\end{overpic} 
\begin{overpic}[width=0.32\textwidth,grid=false,tics=10]{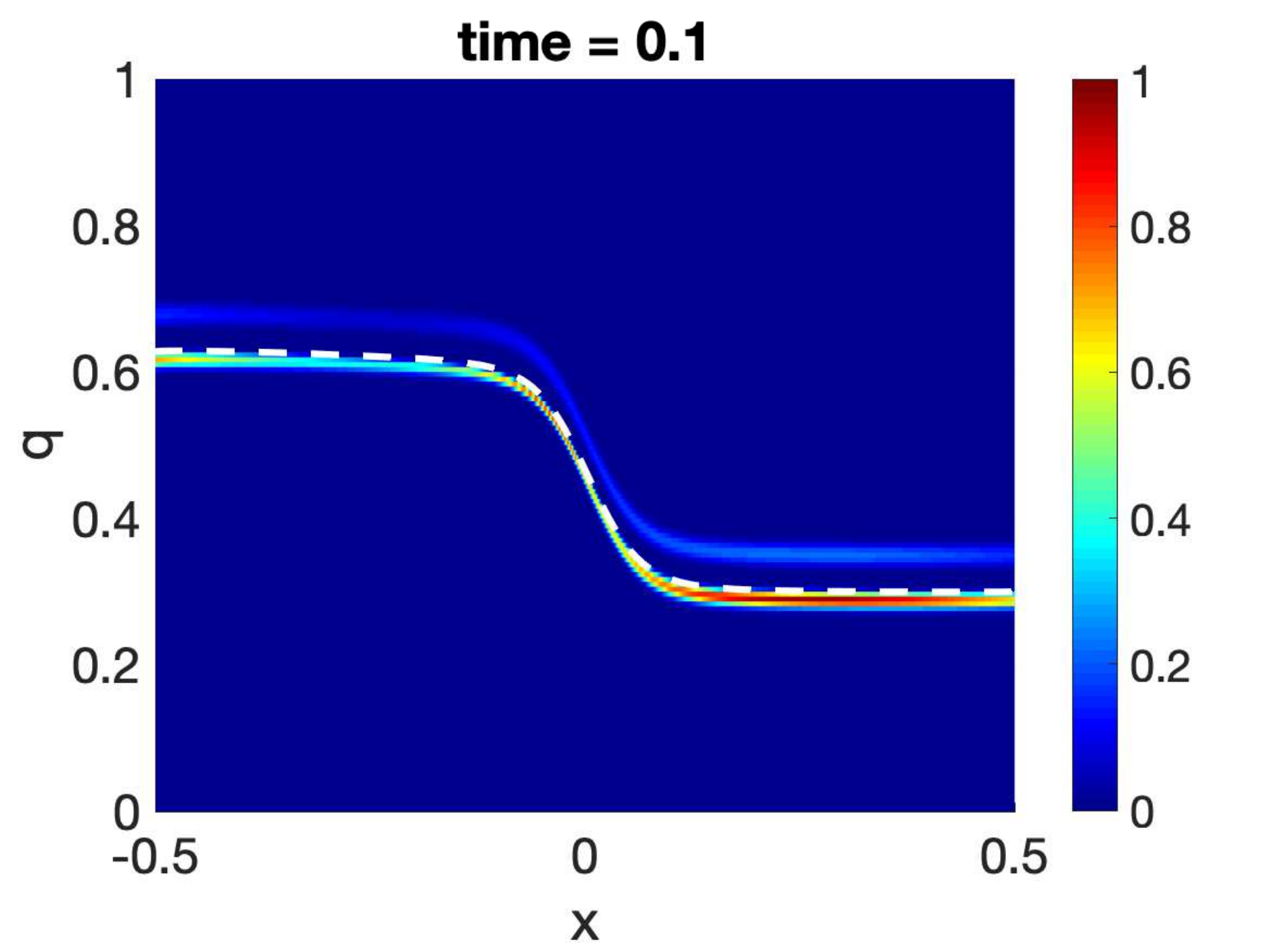}
\put(-10,76){\quad \, $\gamma=100$, $R=0.1$ ($\Delta Q < 2\gamma R$)}
\end{overpic} 
\begin{overpic}[width=0.32\textwidth,grid=false,tics=10]{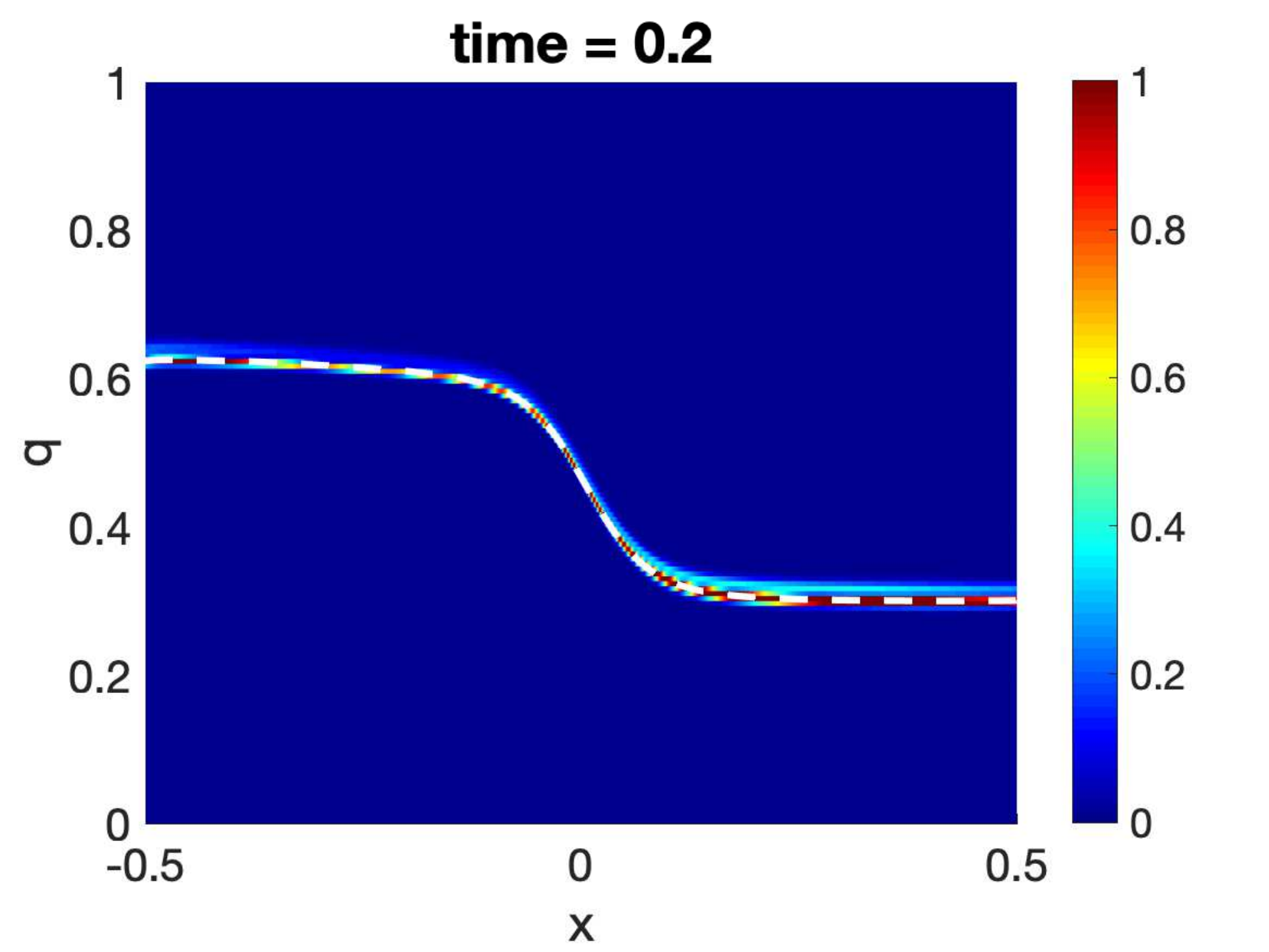}
\end{overpic} 
\caption{Computed distribution density at times $t= 0.05$ s (left), $t= 0.1$ s (center), $t = 0.2$ s (right)
for the case of $v = q$ with $\gamma = 100$ and different values of $R$: $R = 0.01$ (top),
$R = 0.02$ (center), $R = 0.1$ (bottom). The white dash line represents $ q^\ast(t,x)$.
}\label{R_varies}
\end{figure}

Next, we consider the case of $v = q$ and $\Delta Q \leq 2 \gamma R$, i.e.~people's paths cannot cross if the hypotheses
of Theorem 1 are satisfied. We consider the same initial condition as for the results in Fig.~\ref{Contagion_ref} (a) and (b) 
and keep $\gamma$ constant ($\gamma = 100$), 
but vary the value of $R$. Fig.~\ref{R_varies} reports the computed distribution density at 
three times for three different values of $R$, with $\Delta Q = 2 \gamma R$ (top) and $\Delta Q < 2 \gamma R$ (center and bottom).
Like the results in Fig.~\ref{Contagion_ref}, as time passes $h(x,q)$ starts to concentrate on $q^\ast(t,x)$. In addition, 
as $R$ gets bigger, and thus $2 \gamma R$ becomes larger than $\Delta Q$, the slope of $q^\ast$ around $x = 0$ gets milder.

\section{Coupling the pedestrian dynamics model to the contagion model} \label{sec:coupling}

In this section, we couple the 1D pedestrian dynamics model \eqref{eq:1dmodel}
to the contagion model \eqref{eq:disease_kineticeq} to simulate the onset of a disease spreading in a confined
environment. For such coupling, we need to rewrite the contagion model to account for the
fact that we have groups of people with different walking directions.
Thus, we introduce $h^i(t,x,q)$, which is the probability of finding people with contagion level 
$q$ at time $t$, position $x$, and with walking direction $\theta_i$. Following \eqref{eq:disease_kineticeq}, we have:
\begin{equation} \label{eq:disease_kineticeq_dir} 
h^i_t+(v^ih^i)_x=\gamma((q-q^{i, \ast})h^i)_q,
\end{equation}
where quantity $q^{i, \ast}(t,x)$ is the local \textit{average} contagion level of infectious disease weighted by the distance to $x$:
\begin{equation} \label{q_act}
\quad q^{i, \ast}(t,x)= \frac{\int \int \kappa(|x-y|) h^i(t,y,q)qdqdy}{\int \int \kappa(|x-y|)h^i(t,y,q)dqdy}.
\end{equation}

The couple model reads: find $f^i$ and $h^i$ such that
\begin{empheq} [left=\empheqlbrace]{align}
&\dfrac{\partial f^i}{\partial t} + \dfrac{\partial \left( v^i [\rho] (t, x) f^i(t, x) \right)}{\partial x} = \mathcal{J}^i[f](t, x), \label{eq:coupledpedestrians} \\
&\dfrac{\partial h^i}{\partial t}+\dfrac{\partial (v^i h^i)}{\partial x}=\gamma\dfrac{\partial ((q-q^{i, \ast})h^i)}{\partial q}.  \label{eq:contagion} 
\end{empheq}
for $i=1,2$. Note that system \eqref{eq:coupledpedestrians}-\eqref{eq:contagion} is one way coupled: 
eq.~\eqref{eq:contagion} depends on the solution to eq.~\eqref{eq:coupledpedestrians} through \eqref{eq:rho}-\eqref{eq:v^i}, while eq.~\eqref{eq:coupledpedestrians} does not depend on the solution to 
eq.~\eqref{eq:contagion}.

\subsection{Numerical method}\label{sec:coupled_method}

The one-way coupling of system \eqref{eq:coupledpedestrians}-\eqref{eq:contagion} simplifies
the numerical algorithm to approximate its solution. At every time level $t^{k+1}$, we first compute
distribution function $f^{i,k+1}$, with which we can compute $\rho^{k+1}$ and  $v^{i,k+1}$.
Then, we use $v^{i,k+1}$ to compute contagion level $h^{i,k+1}$. 

To be more specific, the algorithm is as follows.
Given initial conditions $f^{i,0}=f^i(0, x)$ and $h^{i,0}=h^i(0,x)$, for $i=1,2$, perform the following steps
for $k=0,1,2, \dots, N_t-2$:

\begin{enumerate}
\item[1a.] Find $f^i$, for $i = 1, 2$, such that problem \eqref{eq:1dstep1} is satisfied. 
 Set $f^{i,k+\frac{1}{2}}=f^i(t^{k+1}, x)$. 
 
\item[1b.] Find $f^i$, for $i = 1, 2$, such that problem \eqref{eq:1dstep3} is satisfied. 
Set $f^{i,k+1}=f^i(t^{k+1}, x)$. Then, compute $\rho^{k+1}$ from \eqref{eq:rho}
and $v^{i,k+1}$ from \eqref{eq:v}-\eqref{eq:v^i}.
\end{enumerate}

\begin{enumerate}
\item[2.] 
Find $h^i$, for $i = 1, 2$, such that\\
\begin{equation}\label{eq:disease_chap7}
\begin{cases}
\dfrac{\partial h^i}{\partial t}+\dfrac{\partial (v^{i,k+1} h^i)}{\partial x}=\gamma \dfrac{\partial ((q-q^{i, \ast})h^i)}{\partial q} \,\,\, \text{on } (t^k, t^{k+1})  \\ 
h^i(t^{k}, x)=h^{i, {k}
}\end{cases}
\end{equation}
 Set $h^{i,k+1}=h^i(t^{k+1}, x)$ and $h^{k+1} = \sum_{i = 1}^2 h^{i,k+1}$.
\end{enumerate}

For the full discretization of the problems at step 1a and 1b we use the schemes described in 
Sec.~\ref{sec:1d-discretization}, while for the full discretization of the problem at step 2 we use
the method reported in Sec.~\ref{sec:contagion_full_disc}.

\subsection{Numerical results}
We test the approach presented in Sec.~\ref{sec:coupled_method} on three 1D problems, corresponding
to unidirectional or bidirectional pedestrian flow in a narrow corridor. For all the problems, 
the computational domain in the $xq$-plane is $[-15, 15] \times [0, 1]$ and it is occupied by a group of 
38 pedestrians. 
In test 1 there is only one exit at $x = 15$, while in test 2 and 3 there is an exit at each end of the corridor.
For all the tests, we set $\varepsilon=0.4$, $\alpha=1$, $\gamma = 100$ and $R = 0.01$. The dimensionless quantities are obtained 
by using the following reference quantities: $D=30$ m, $v_M= 2$ m/s, $T = 15$ s, $\rho_M = 3$ people/$m$, $q_{max} = 1$.

We use different mesh sizes for the pedestrian dynamics model and for the contagion model in order to ensure that we have the right
level of refinement for both models and, at the same time, contain the computational cost. We consider mesh size $\Delta x_k=0.05$ m
for the pedestrian dynamics and $\Delta x_d=0.025$ for the disease contagion model. The associated time steps
are chosen to satisfy the CFL conditions: $\Delta t_k=0.003$ s and  $\Delta t_d=0.00002$ s for the pedestrian dynamics
and contagion models, respectively.

\begin{figure}
\begin{center}
\begin{overpic}[width=0.33\textwidth,grid=false,tics=10]{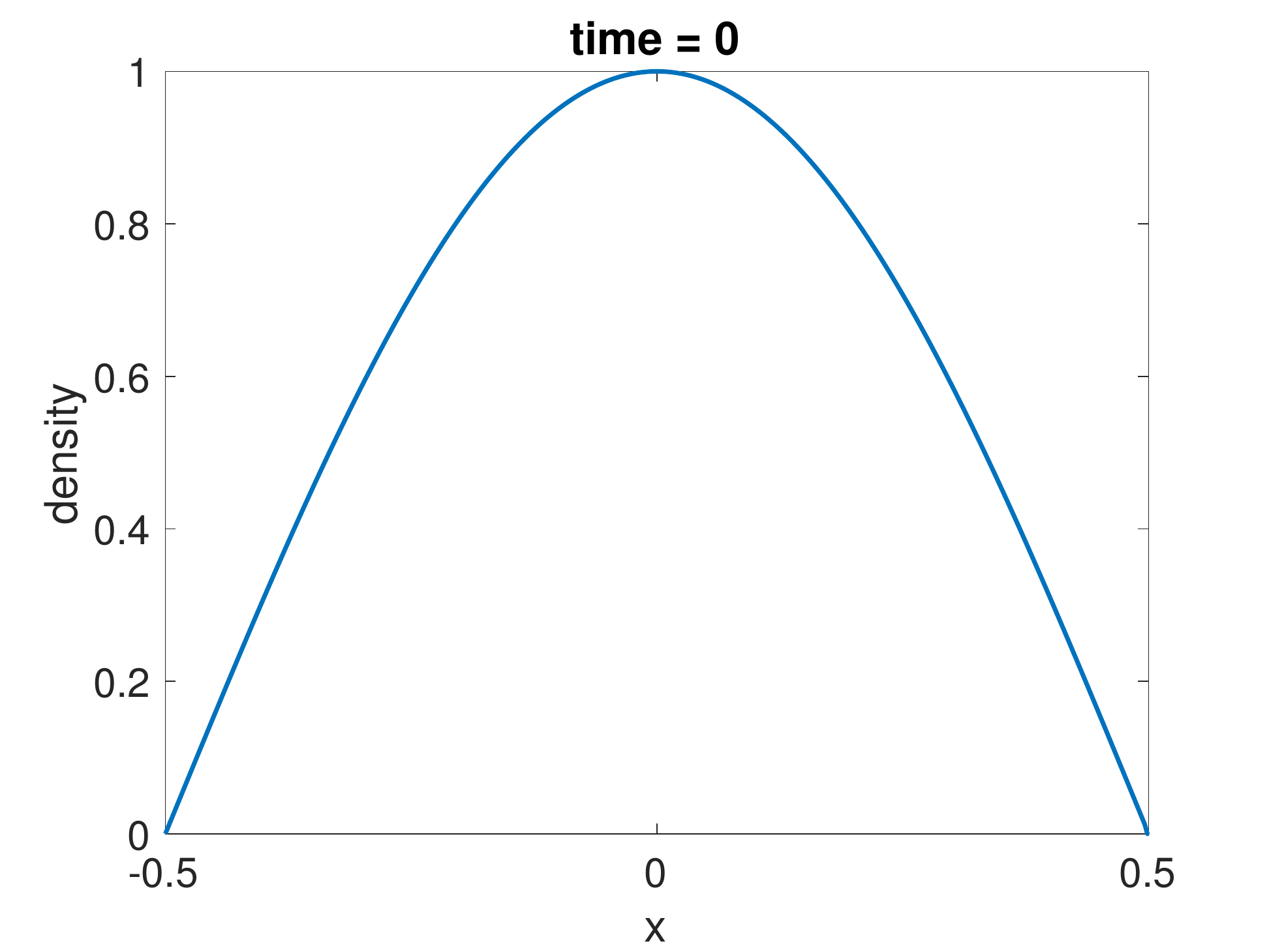}
\end{overpic}
\begin{overpic}[width=0.33\textwidth,grid=false,tics=10]{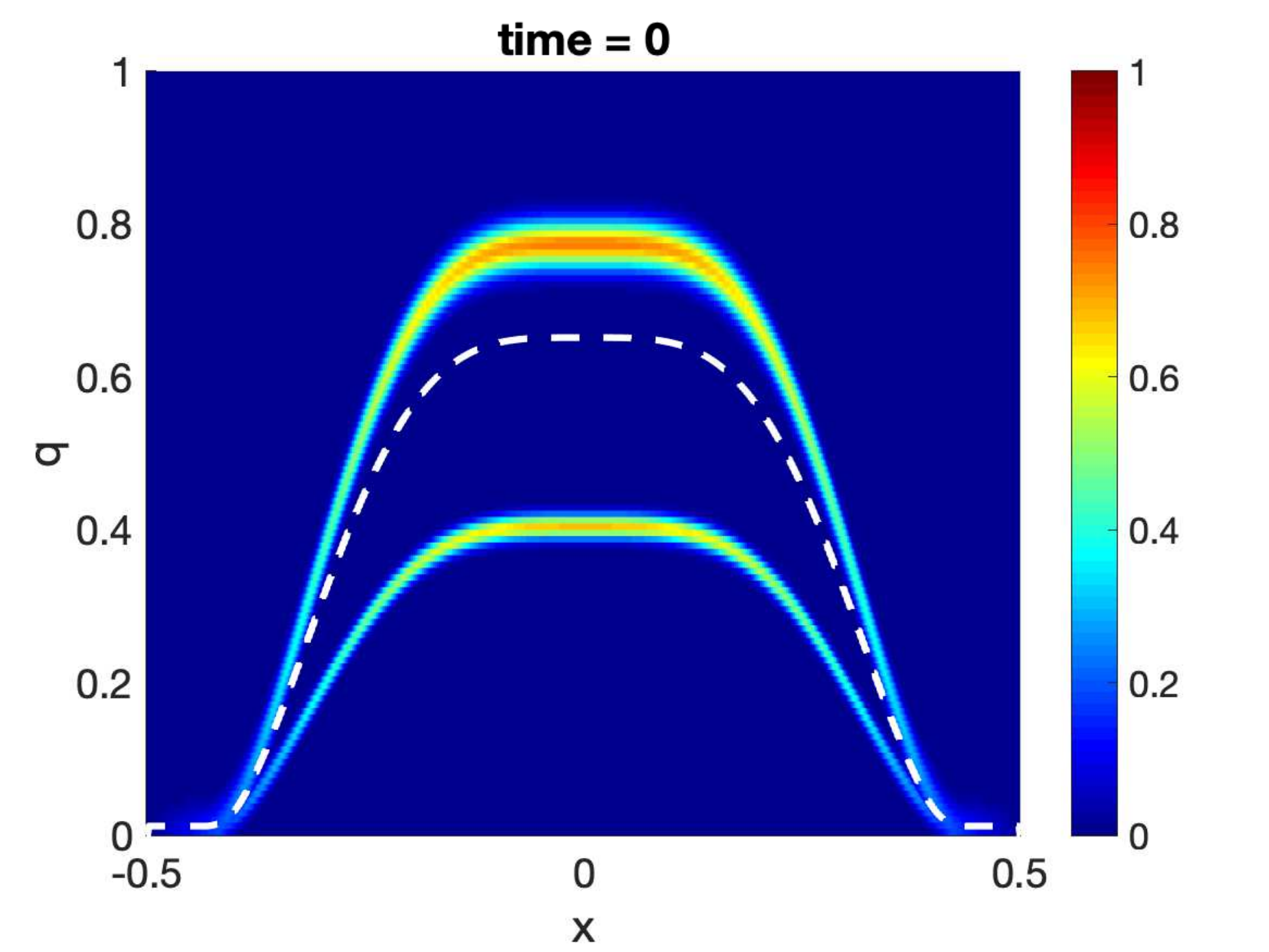} 
\end{overpic} 
\begin{overpic}[width=0.33\textwidth,grid=false,tics=10]{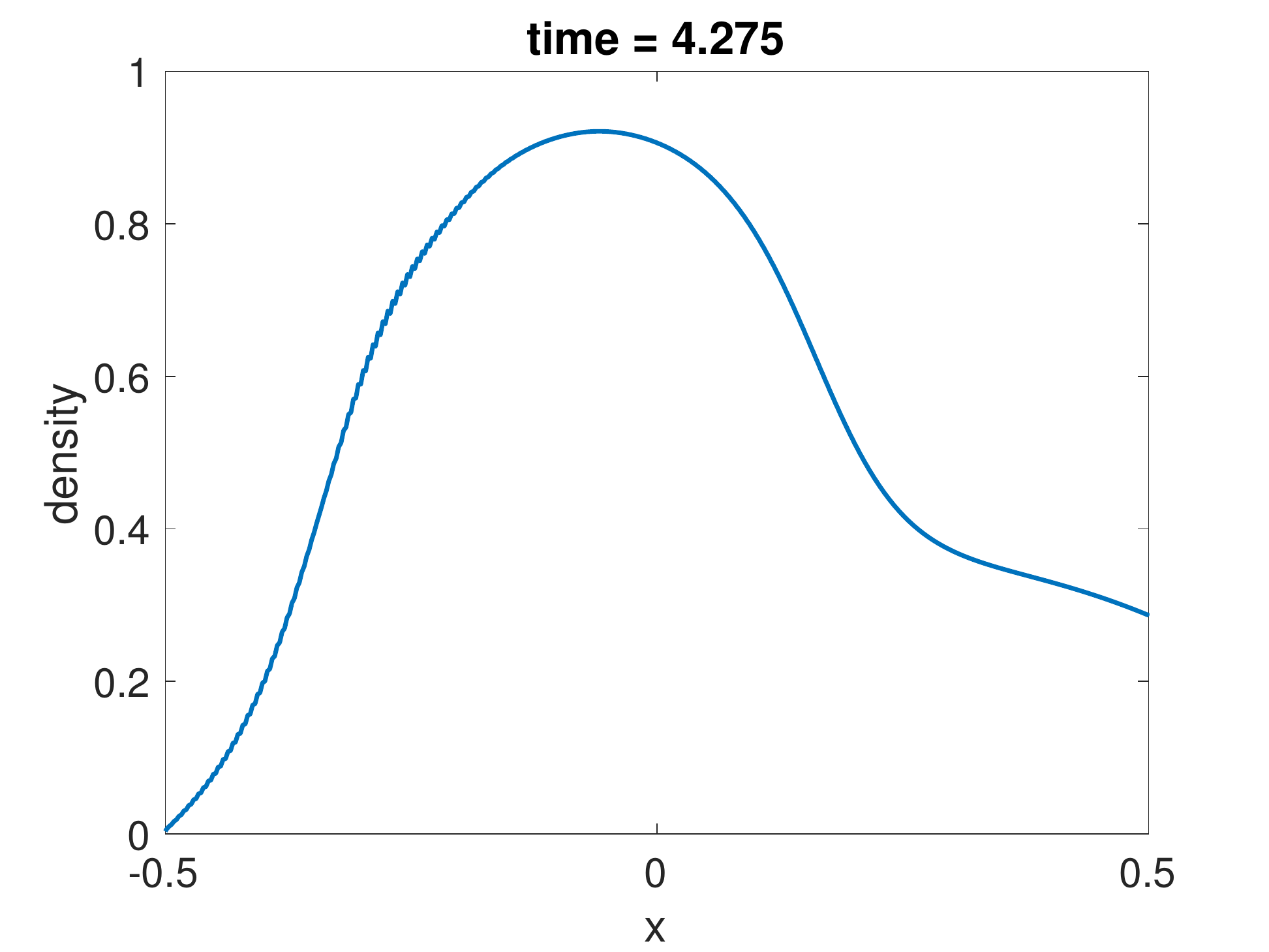}
\end{overpic}
\begin{overpic}[width=0.33\textwidth,grid=false,tics=10]{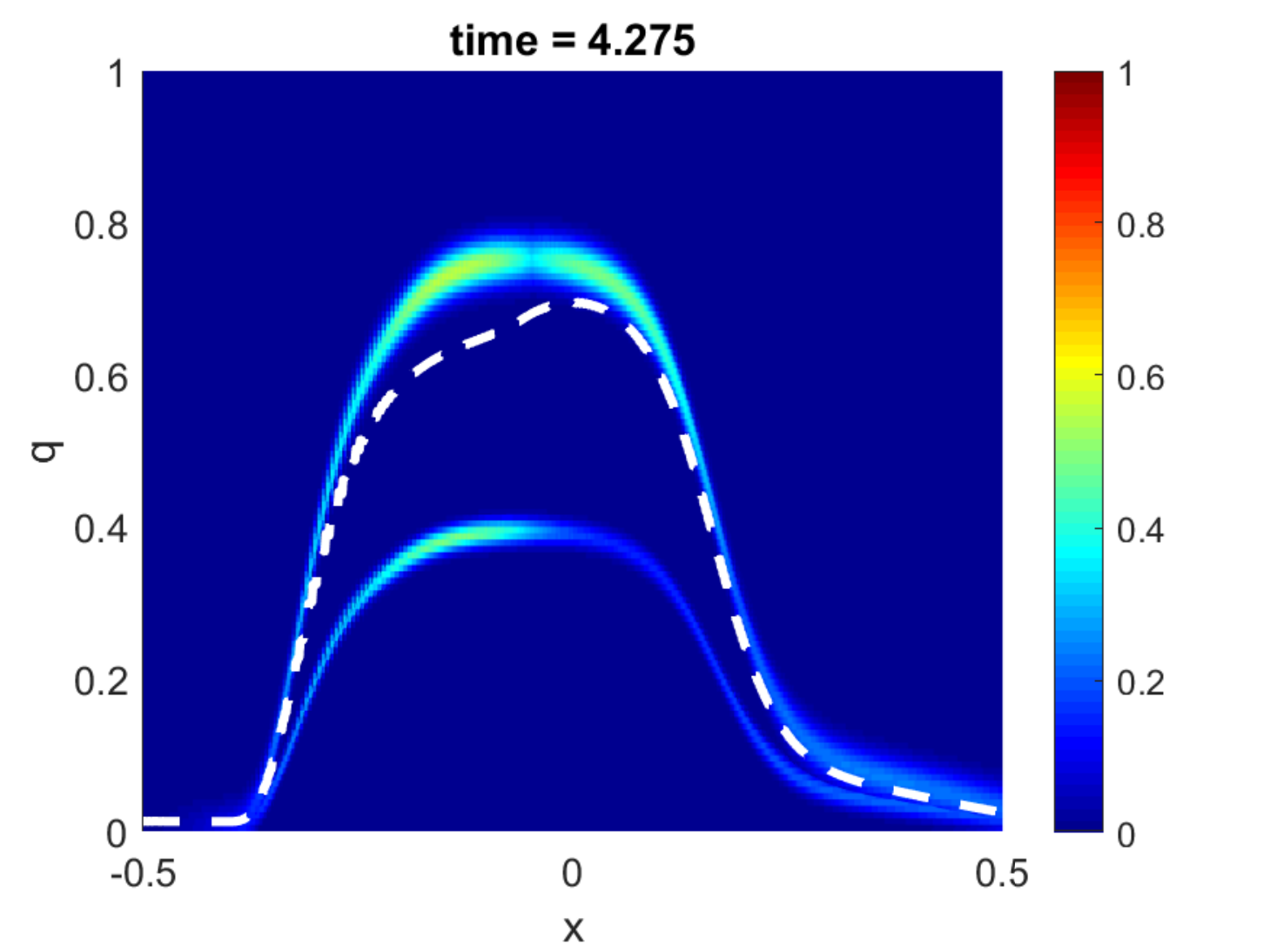} 
\end{overpic} \\
\begin{overpic}[width=0.33\textwidth,grid=false,tics=10]{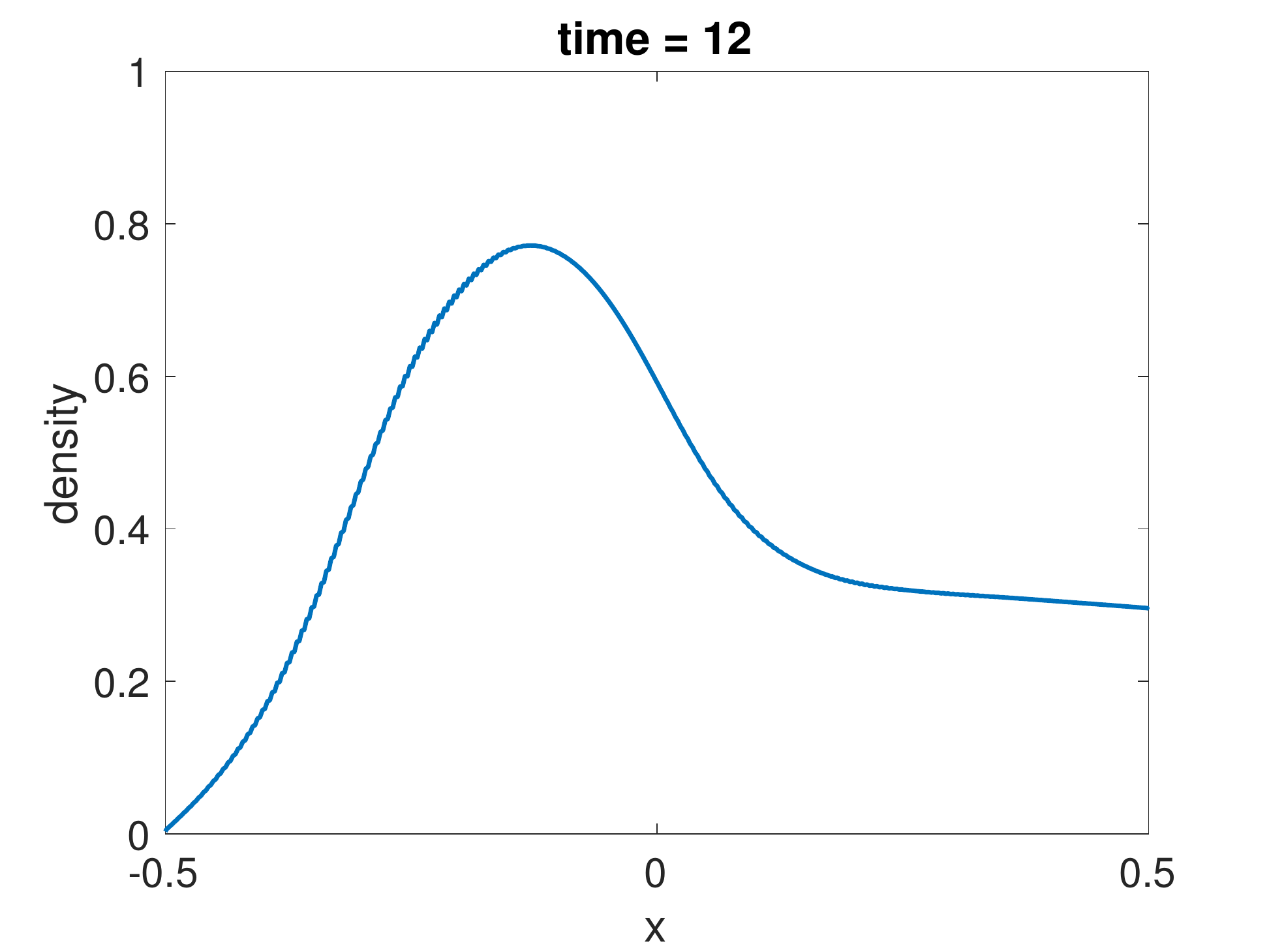}
\end{overpic}
\begin{overpic}[width=0.33\textwidth,grid=false,tics=10]{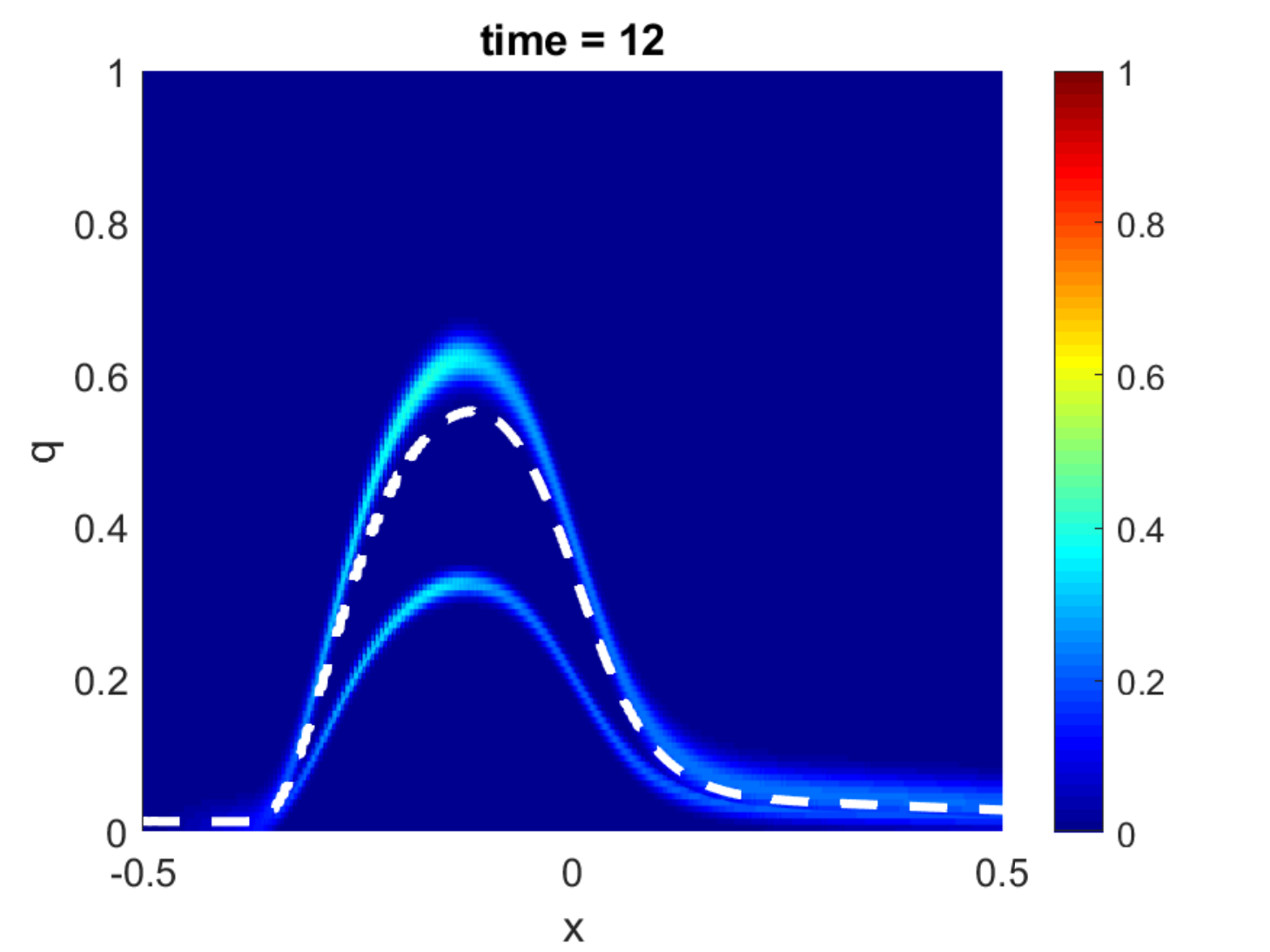} 
\end{overpic} \\
\end{center}
\caption{Test 1: computed people density (left) and
corresponding distribution density $h$ (right) for $t = 0$ s (top), $t = 4.275$ s (middle), and $t = 12$ s (bottom). On the right, the color represents 
the probability of finding sick people.} 
\label{coupledmodel_1}
\end{figure}

{\bf Test 1}. 
The group of 38 pedestrians is initially placed as shown in Fig.~\ref{coupledmodel_1} top left panel, with initial direction $\theta_1$.
This 1D test represents a small crowd walking with the same direction through a corridor that has an exit at its end, i.e.~at $x = 15$ m.
The corresponding initial distribution density $h$ is shown in Fig.~\ref{coupledmodel_1} top right panel. 
Just like for the tests in Sec.~\ref{sec:num_res_contagion}, which are inspired from the work in \cite{Bertozzi2015}, 
we take initial $h$ with two bumps in $q$ for every $x$. Since the corridor is initially rather densely populated towards the center, 
we consider an initial large value of $q$ around $x = 0$.

Fig.~\ref{coupledmodel_1} shows the evacuation process: computed people density (left) and corresponding distribution density $h$
(right).
As expected, throughout the entire time interval there is a higher probability of finding infected people (i.e.~higher values of $h$)
where there is a larger crowd.
As time passes, people move toward the exit and the probability of finding infected people decreases.

\begin{figure}
\begin{center}
\begin{overpic}[width=0.33\textwidth,grid=false,tics=10]{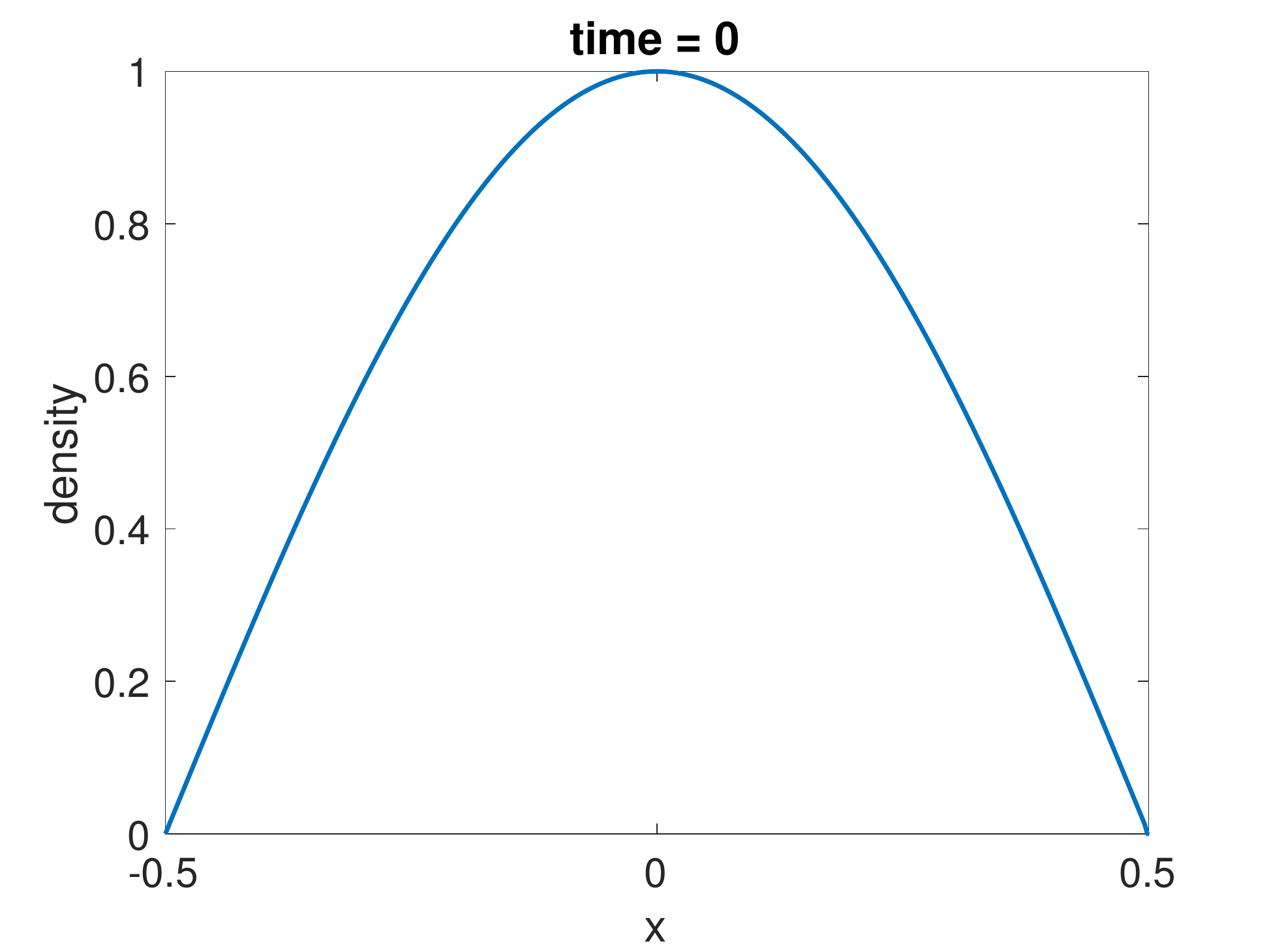}
\end{overpic}
\begin{overpic}[width=0.33\textwidth,grid=false,tics=10]{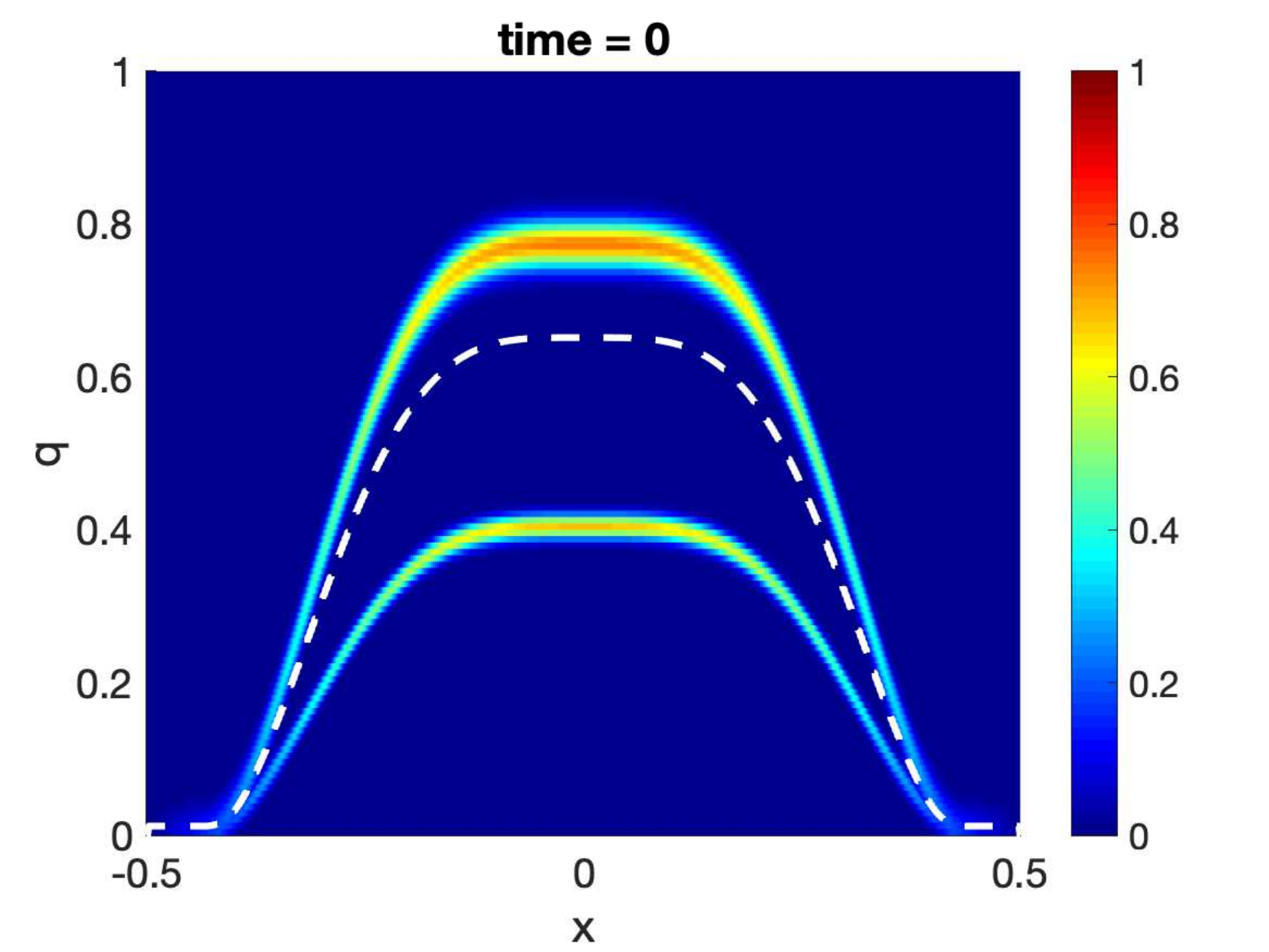} 
\end{overpic} 
\begin{overpic}[width=0.33\textwidth,grid=false,tics=10]{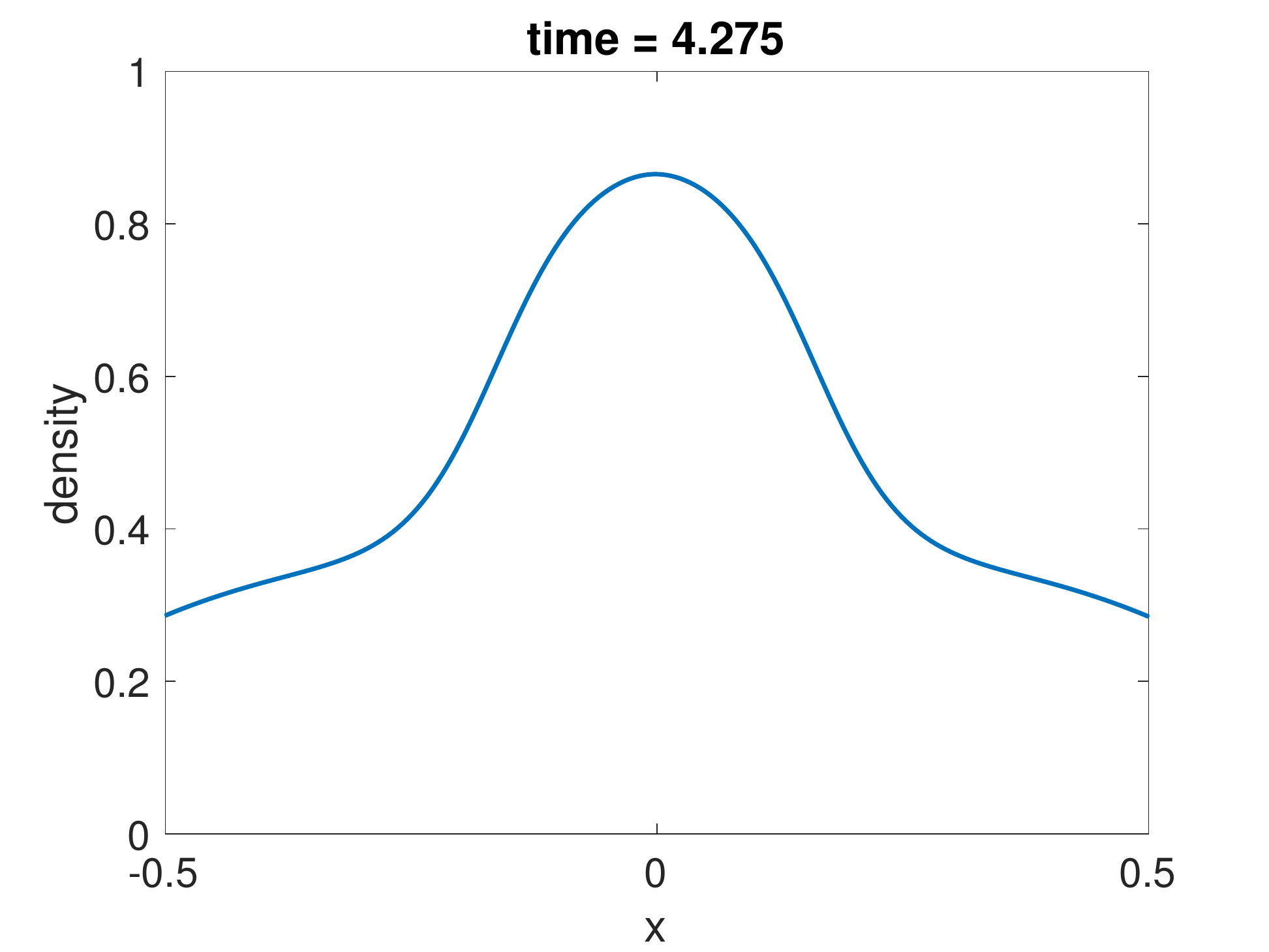}
\end{overpic}
\begin{overpic}[width=0.33\textwidth,grid=false,tics=10]{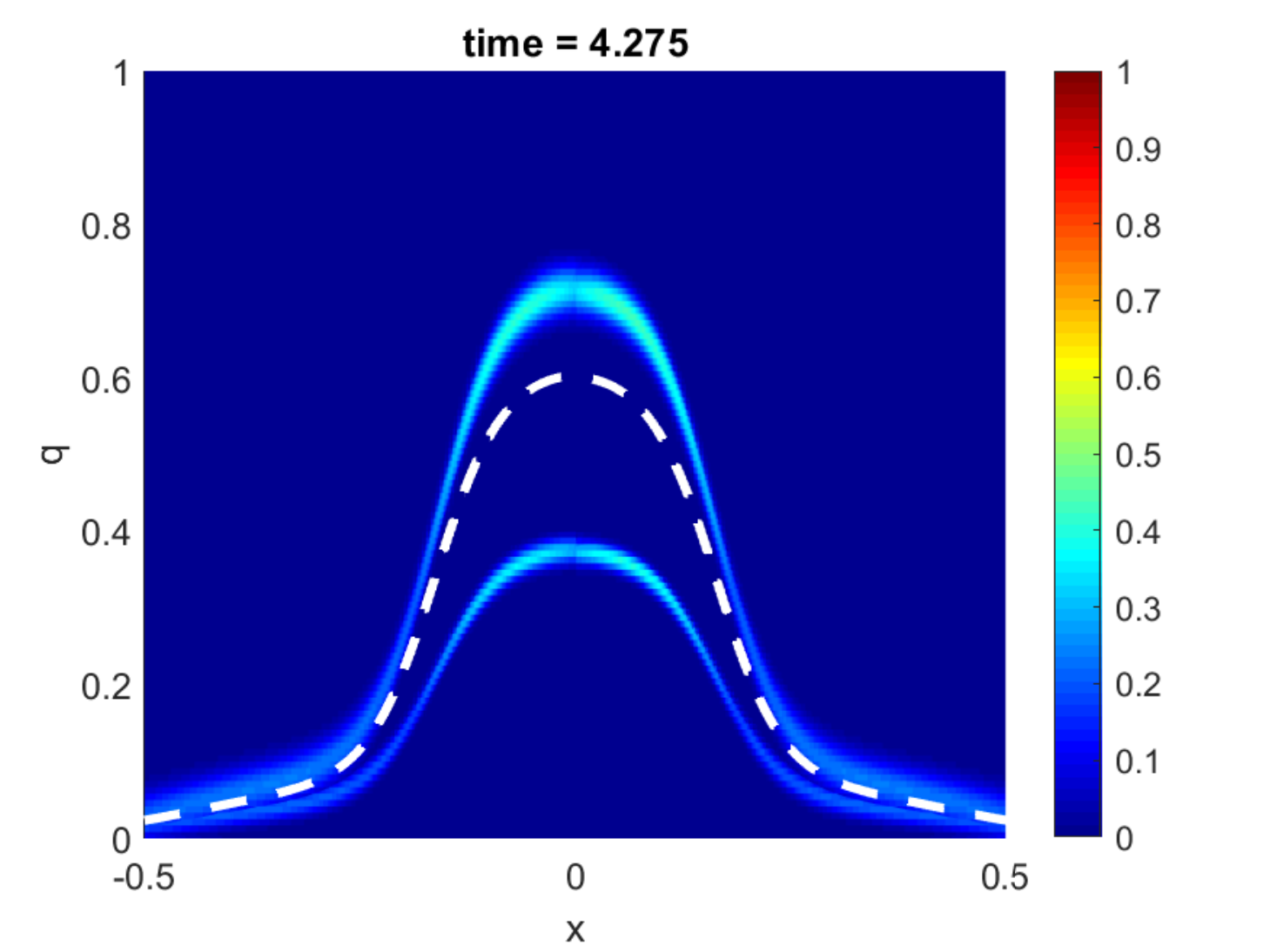} 
\end{overpic} \\
\begin{overpic}[width=0.33\textwidth,grid=false,tics=10]{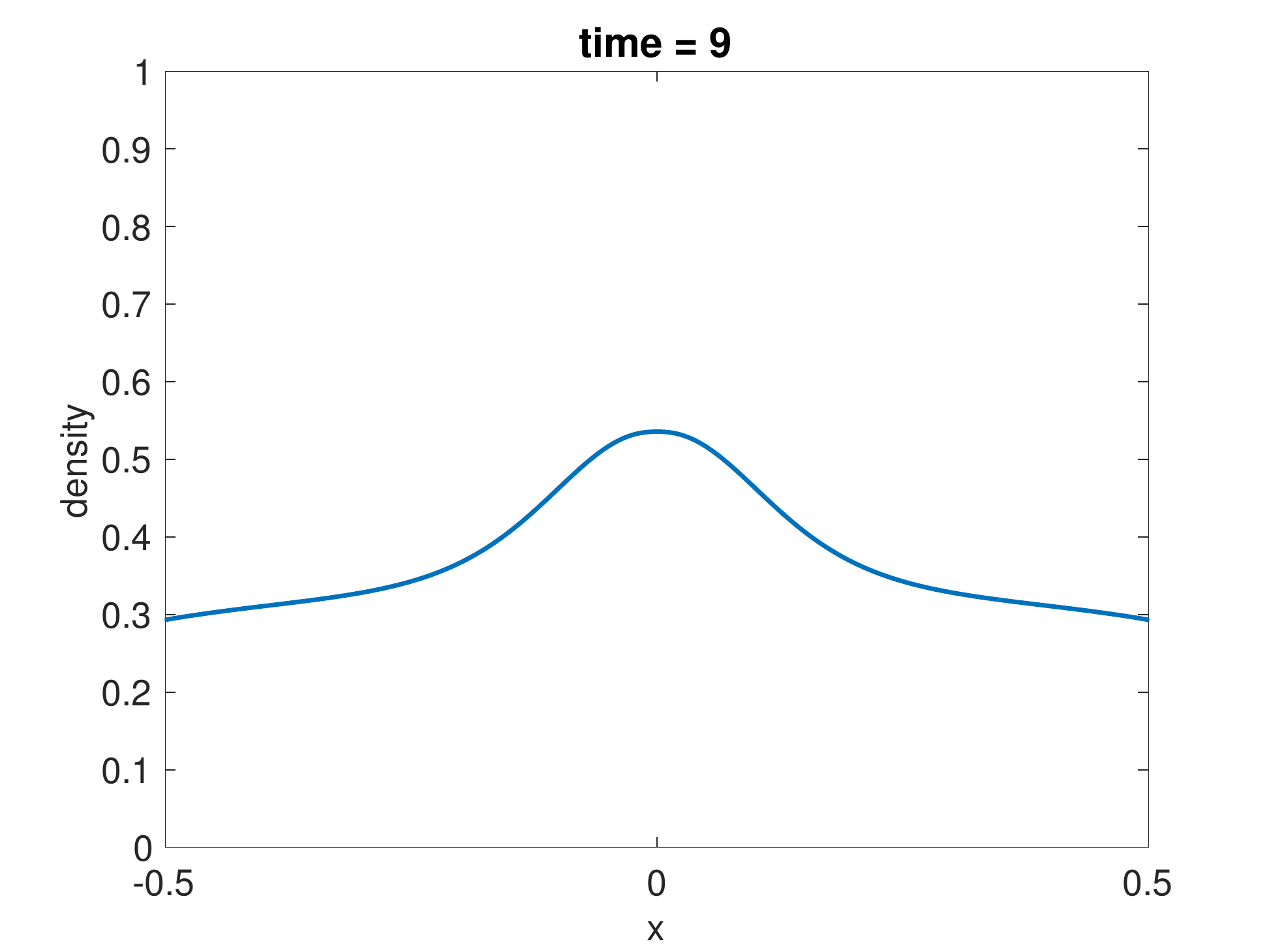}
\end{overpic}
\begin{overpic}[width=0.33\textwidth,grid=false,tics=10]{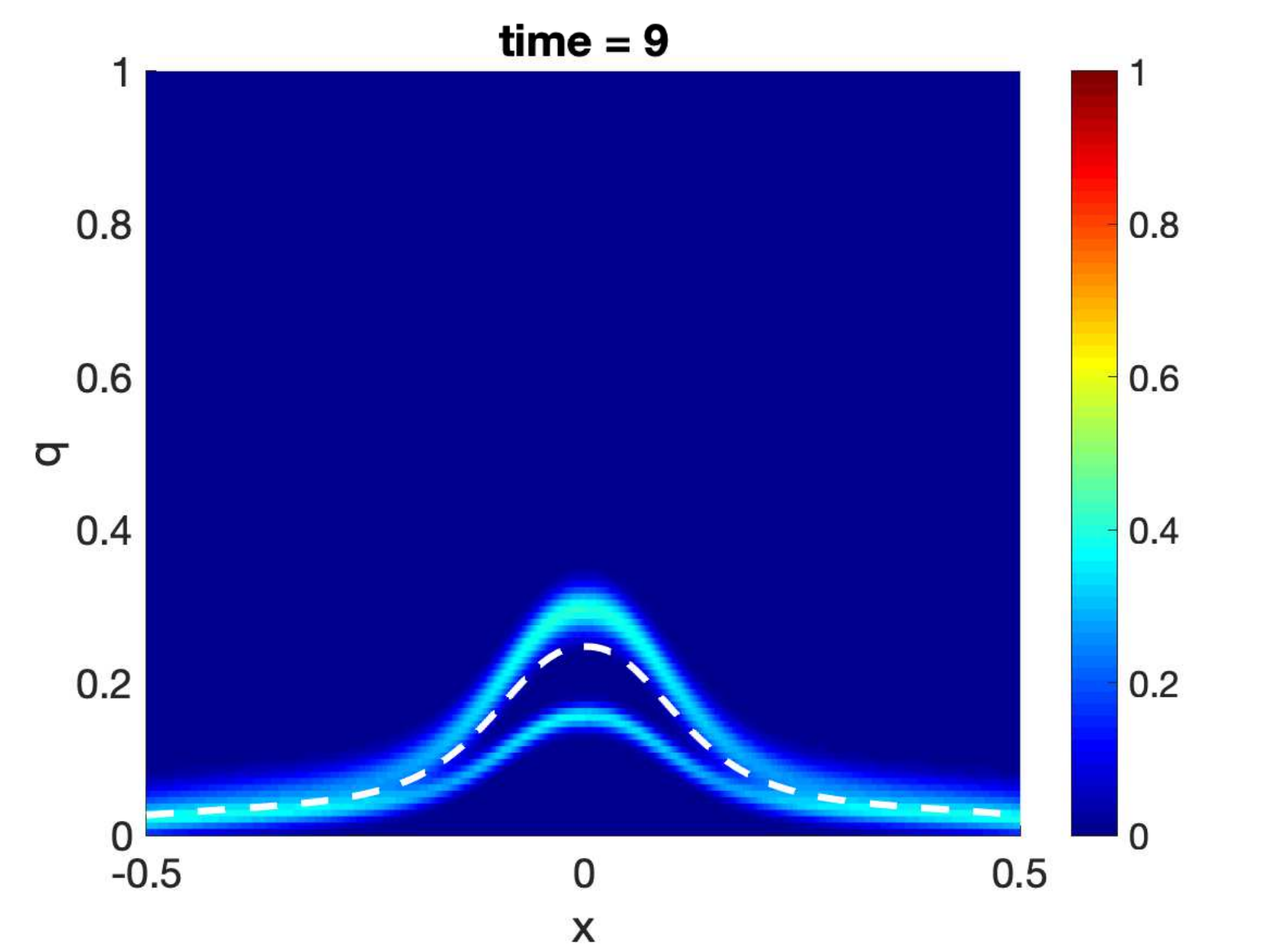} 
\end{overpic} \\
\end{center}
\caption{Test 2: computed people density (left) and
corresponding distribution density $h$ (right) for $t = 0$ s (top), $t = 4.275$ s (middle), and $t = 9$ s (bottom). On the right, the color represents 
the probability of finding sick people.} 
\label{coupledmodel_2}
\end{figure}

{\bf Test 2}. 
The initial positioning of the 38 pedestrians, shown in Fig.~\ref{coupledmodel_2} top left panel, 
is the same as for test 1. 
Moreover, the initial distribution density $h$, which is  shown in Fig.~\ref{coupledmodel_2} top right panel, is the same 
as for test 1.
The difference with respect to test 1 is that there is an exit at each end of the corridor. The people with initial position $x_0 \geq 0$
(resp. $x_0 < 0$) have initial direction $\theta_1$ (resp., $\theta_2$), i.e.~they are headed to the right (resp., left) exit.

Fig.~\ref{coupledmodel_2} shows the evacuation process: computed people density (left) and 
corresponding distribution density $h$ (right). Thanks to the addition of an exit, we see that the probability 
of finding sick people decreases faster than for test 1: compare the middle right panel in Fig.~\ref{coupledmodel_1} 
and \ref{coupledmodel_2}, which both refer to time $t = 4.275$ s. 
This is obviously due to the fact that it takes less time to leave the corridor and thus the crowd density
rapidly decreases. 

\begin{figure}
\begin{center}
\begin{overpic}[width=0.33\textwidth,grid=false,tics=10]{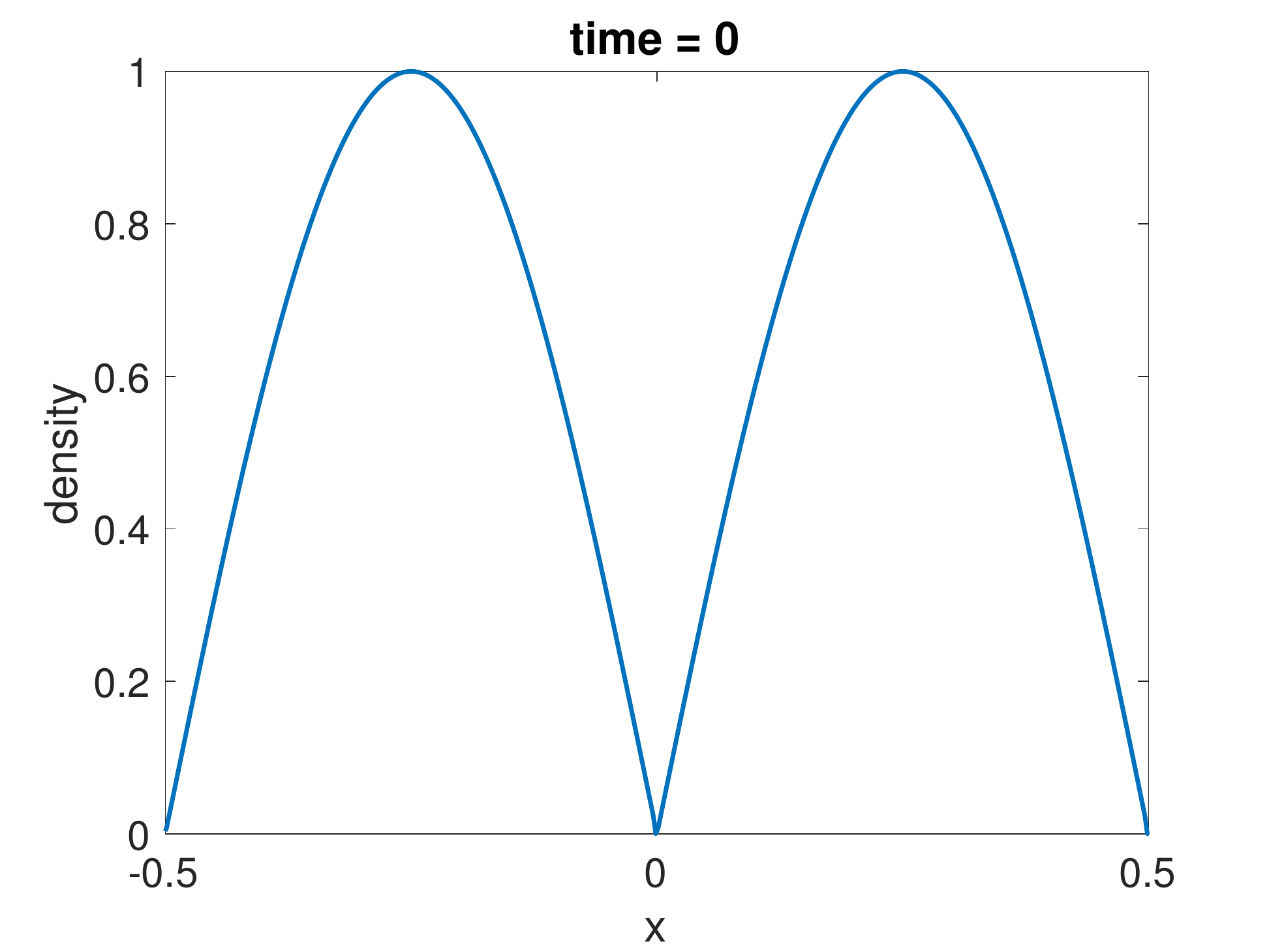}
\end{overpic}
\begin{overpic}[width=0.33\textwidth,grid=false,tics=10]{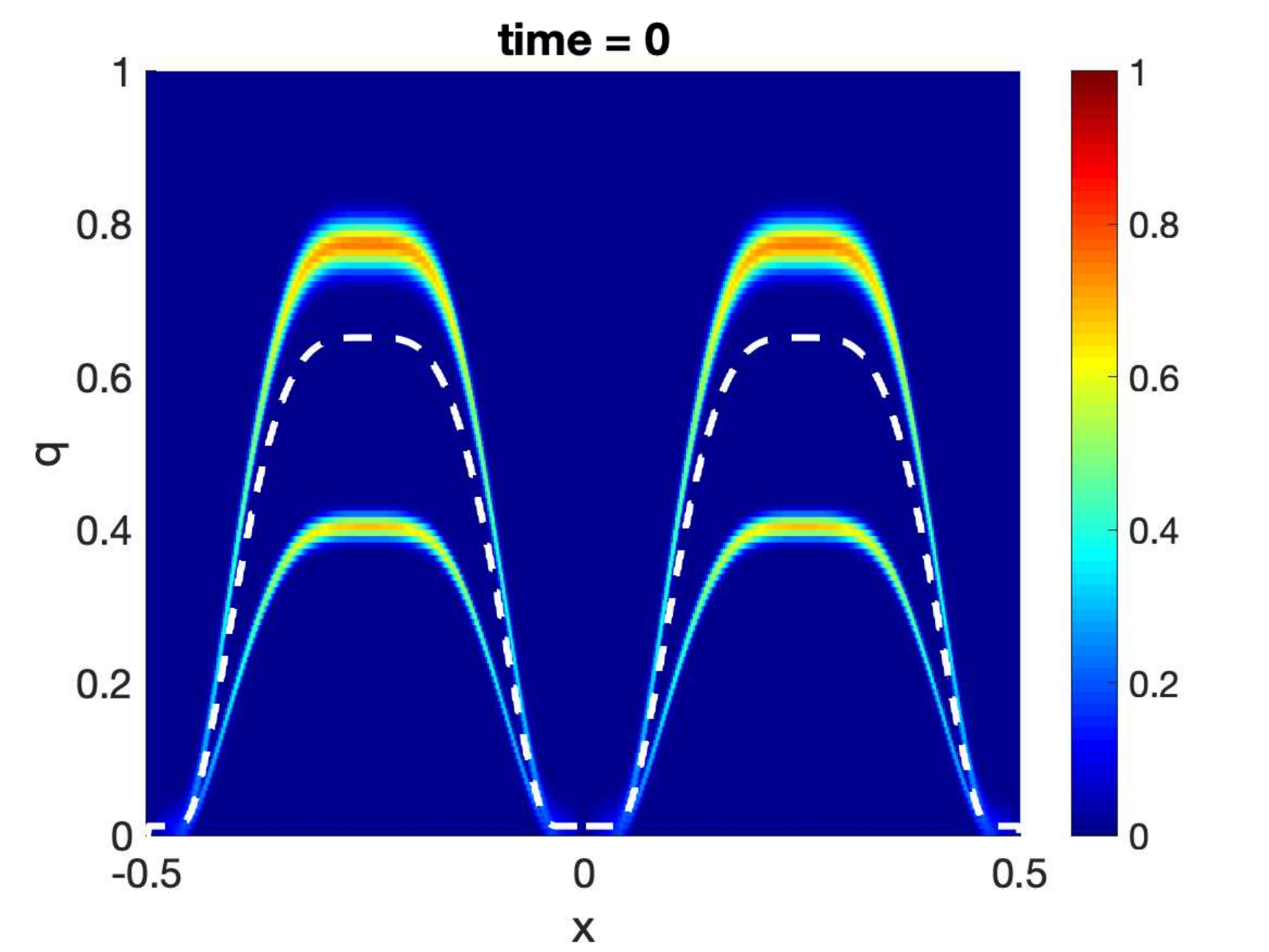} 
\end{overpic} 
\begin{overpic}[width=0.33\textwidth,grid=false,tics=10]{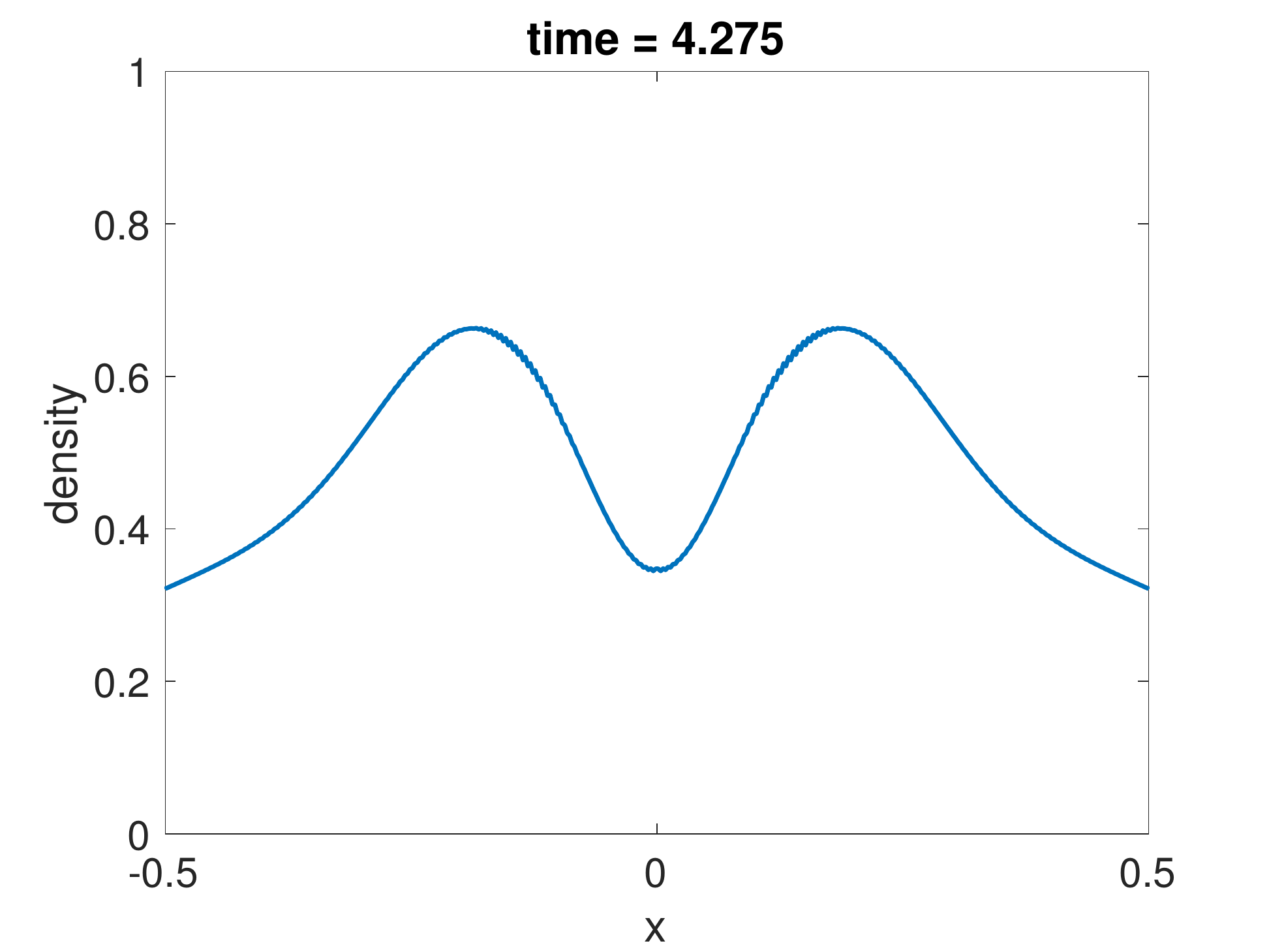}
\end{overpic}
\begin{overpic}[width=0.33\textwidth,grid=false,tics=10]{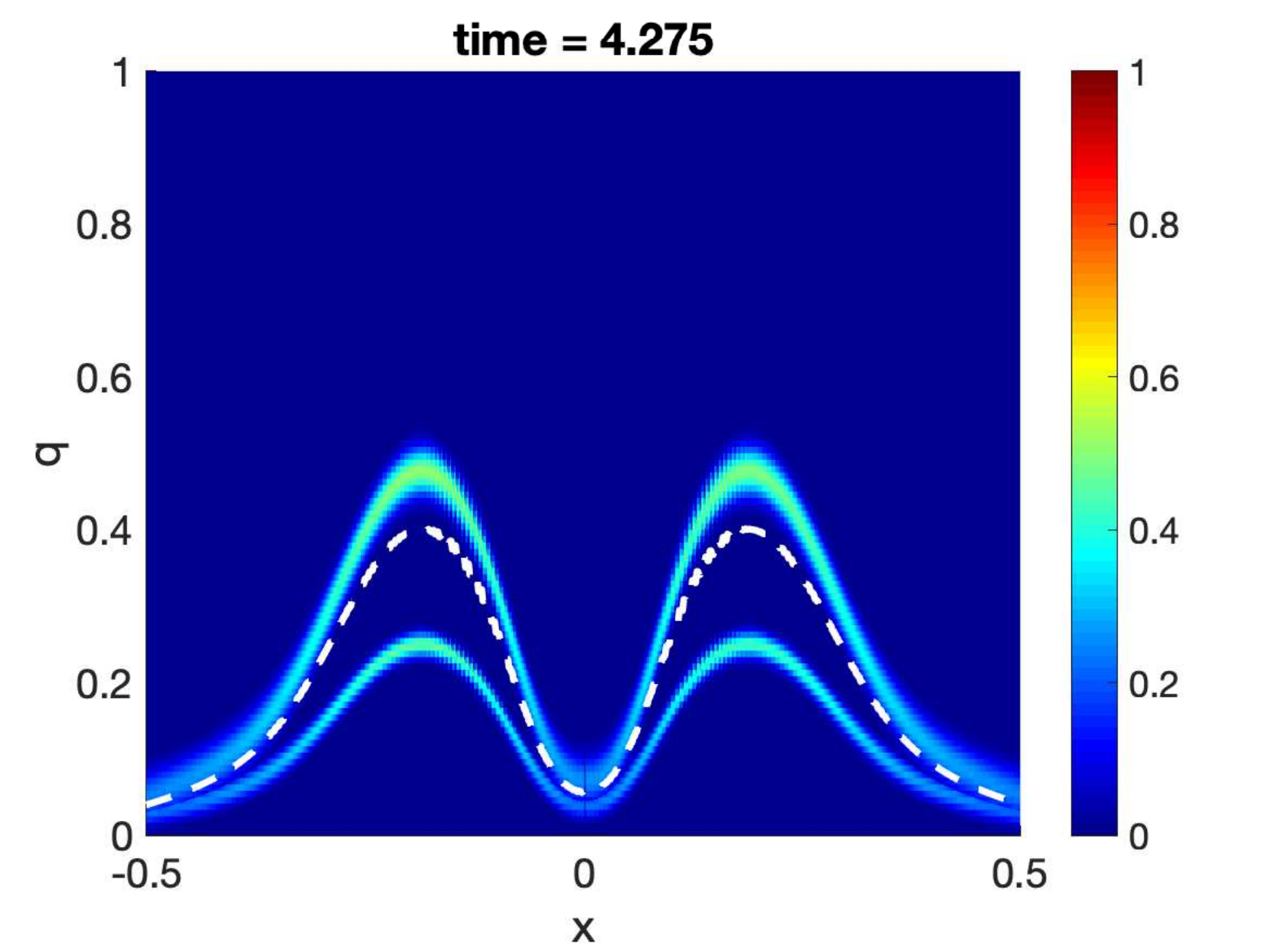} 
\end{overpic} \\
\begin{overpic}[width=0.33\textwidth,grid=false,tics=10]{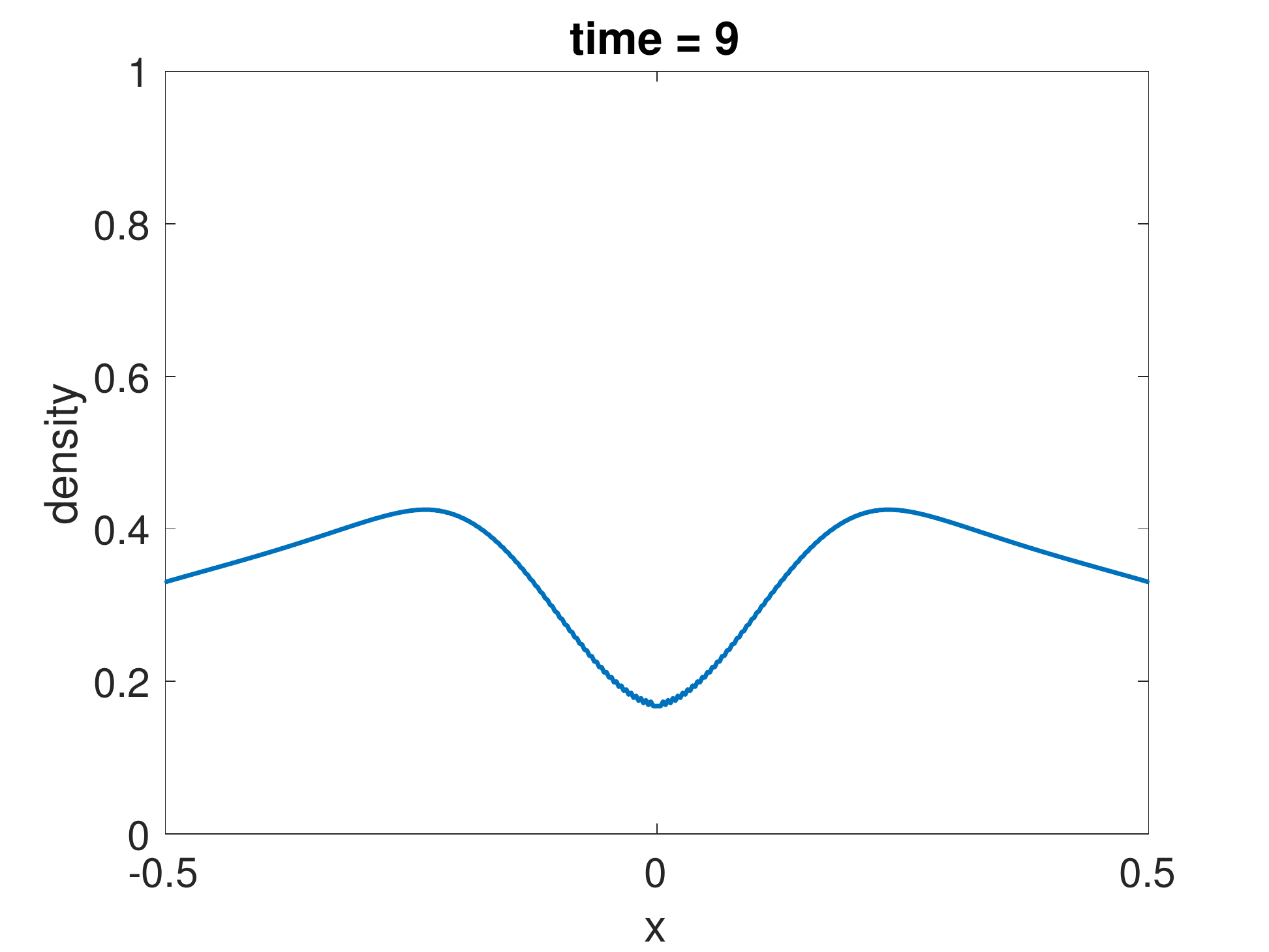}
\end{overpic}
\begin{overpic}[width=0.33\textwidth,grid=false,tics=10]{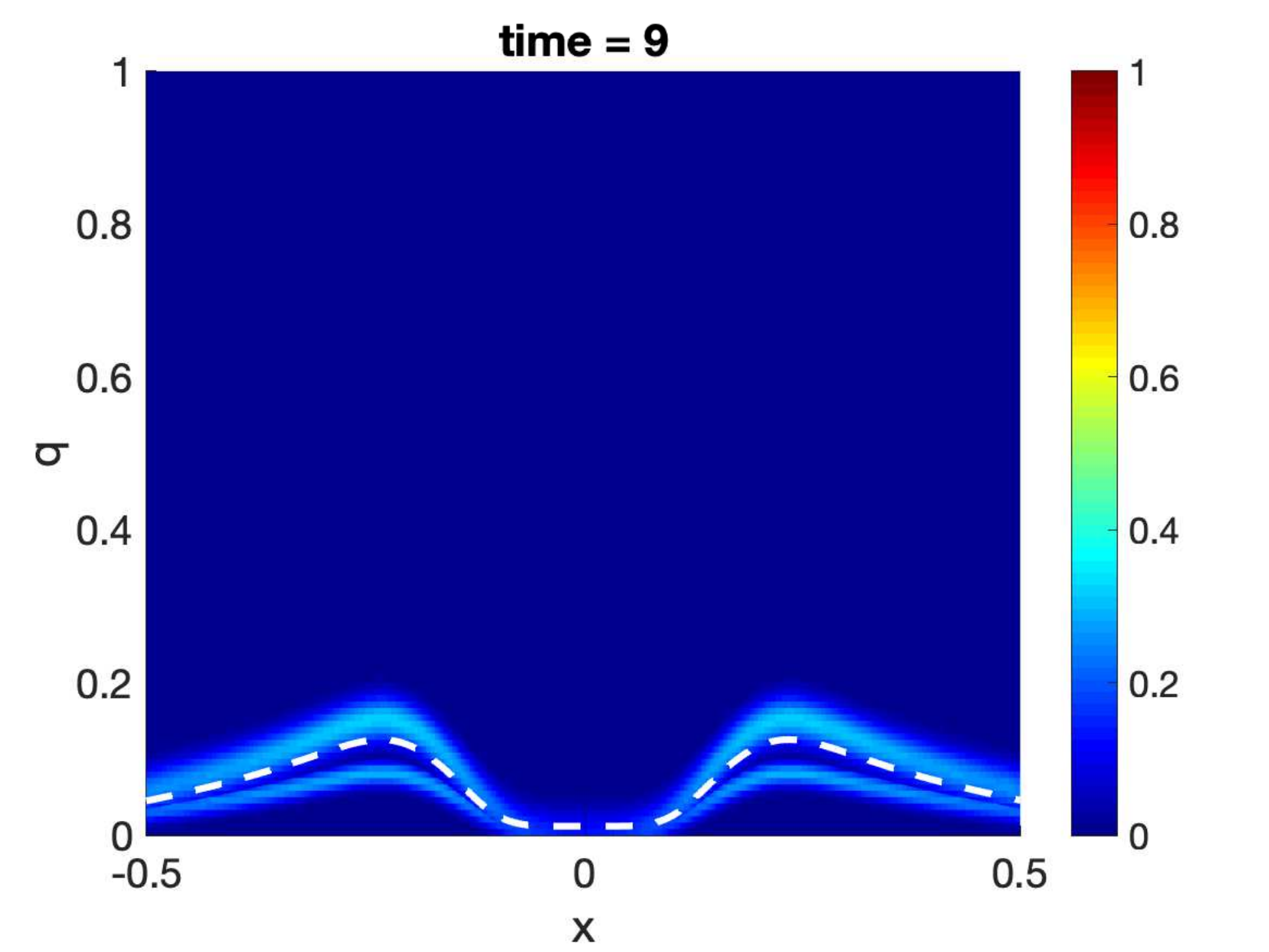} 
\end{overpic} \\
\end{center}
\caption{Test 3: computed people density (left) and
corresponding distribution density $h$ (right) for $t = 0$ s (top), $t = 4.275$ s (middle), and $t = 9$ s (bottom). On the right, the color represents 
the probability of finding sick people.} 
\label{coupledmodel_3}
\end{figure}

{\bf Test 3}. 
The 38 pedestrians are initially placed as shown in Fig.~\ref{coupledmodel_3} top left panel. The crowds size
is the same as for test 1 and 2, but their initial distribution is different. 
Like for test 2, there is an exit at each end of the corridor. Also the initial direction for the pedestrians is assigned as
in test 2: the people with initial position $x_0 \geq 0$ (resp. $x_0 < 0$) have initial direction $\theta_1$ (resp., $\theta_2$).
The initial distribution density $h$ is  shown in Fig.~\ref{coupledmodel_3} top right panel.

Fig.~\ref{coupledmodel_3} shows the evacuation process: computed people density (left) and 
corresponding distribution density $h$ (right). Since the crowd is initially split into two groups, instead of having
one big group as in test 2, the process of evacuation from the corridor is faster. As a result, the probability 
of finding sick people decreases faster than for test 2: compare the middle and bottom right panels in Fig.~\ref{coupledmodel_2} 
and \ref{coupledmodel_3}, which refer to the same times. 
 
Tests 1, 2, and 3 show that in order to reduce of the probability of an infections disease spreading the evacuation 
process from a common space, like a corridor, should be facilitated by, e.g., increasing the number of
exits, and large crowded areas  for sustained period of times should be avoided.

\section{Conclusion}\label{sec:concl}
In order to study the spreading of an infectious disease in a confined environment,
we coupled a kinetic theory approach to model crowd dynamics and a kinetic equation to model contagion.
For the numerical solution of the coupled problem, we
proposed a numerical algorithm that at every time step solves one crowd dynamics problem and one
contagion problem. For the numerical approximation of the solution to the crowd dynamics model, 
we applied the Lie splitting scheme which breaks the problem into a pure advection problem and 
a problem involving the interaction with other pedestrians.
We tested the proposed model and numerical approach on three 1D problems, corresponding
to unidirectional and bidirectional pedestrian flow in a narrow corridor with one or two exits. 
The tests indicated that the probability of an infections disease spreading is reduced when the 
evacuation from the corridor is facilitated and when large crowded areas for sustained period of times are
avoided. 

\section*{Acknowledgements}
This work has been partially supported by NSF through grant DMS-1620384.

\bibliographystyle{plain}
\bibliography{contagion_model}

\end{document}